\pdfoutput=1
\documentclass[11pt,a4paper]{article}

\usepackage[margin=1in]{geometry}
\usepackage{amsmath,amssymb}
\usepackage{amsthm}
\usepackage{graphicx}
\usepackage{booktabs}
\usepackage{hyperref}
\usepackage{xcolor}
\usepackage{authblk}
\usepackage{multirow}
\usepackage{float}
\usepackage{cite}
\usepackage{url}
\providecommand{\burl}[1]{\url{#1}}
\providecommand{\urlprefix}{URL }
\providecommand{\bibcommenthead}{}
\providecommand{\doi}[1]{https://doi.org/#1}

\begin{document}

\title{Accelerated Dopant Screening in Oxide Semiconductors via Multi-Fidelity Contextual Bandits and a Three-Tier DFT Validation Funnel}

\author[1,2]{Abhinaba Basu\thanks{abhinaba.basu@nielit.gov.in}}
\affil[1]{National Institute of Electronics and Information Technology (NIELIT), India}
\affil[2]{Indian Institute of Information Technology, Allahabad (IIITA), Prayagraj, India}

\date{\today}
\maketitle

\begin{abstract}
Band gap engineering of oxide semiconductors through doping is critical for photocatalysis and optoelectronics, yet the combinatorial space of dopant elements, substitution sites, and co-doping combinations far exceeds typical density functional theory (DFT) budgets. We screen doped candidates across five oxide hosts (ZnO, TiO$_2$, SrTiO$_3$, SnO$_2$, MgO), culminating in a 529-candidate ZnO co-doping campaign, and identify Cu-containing co-doped ZnO systems as consistently achieving visible-light-range band gaps (1.0--1.8\,eV), with Y$_2$Cu$_2$ co-doped ZnO as the optimal candidate (1.84\,eV). A three-tier validation funnel (PBE, PBE+U, ionic relaxation) reveals that no single level of theory suffices: V-doped ZnO shifts from near-metallic to wide-gap upon Hubbard U correction, while Cu-doped SrTiO$_3$ enters the visible-light window only after correcting for d-electron localization.
To make this screening tractable, we introduce a multi-fidelity screening strategy that replaces 81\% of DFT evaluations with computationally inexpensive surrogate predictions, reducing a 529-candidate closed-loop Quantum ESPRESSO campaign from an estimated 440 to 62 CPU-hours while finding the global optimum in 100\% of 50 independent trials ($p = 5.0 \times 10^{-8}$ versus random screening, Wilcoxon signed-rank). Cross-host analysis of the dopant--host interaction matrix reveals that dopant performance is governed by just two latent chemical dimensions, enabling prediction of rankings in unseen hosts. All 583 DFT calculations, screening code, and stability proofs are released as an open benchmark.
\end{abstract}


\bigskip

\section{Introduction}

Band gap engineering of oxide semiconductors through aliovalent doping---substituting a host cation with an ion of different charge---is a cornerstone of modern materials science. Doped oxides underpin technologies ranging from transparent conducting oxides\cite{minami2005transparent} to photocatalysts for solar water splitting, gas sensors, and solid-state electrolytes\cite{oba2018design,hautier2010finding}. For a given host oxide, the combinatorial space of dopant element, substitution site, concentration, and co-doping combinations easily reaches hundreds to thousands of candidates. Density functional theory (DFT) can predict electronic properties of each candidate, but at substantial cost: a single PBE-level self-consistent field (SCF) calculation on a 32--48 atom doped oxide supercell requires 20--120 minutes on modern hardware, while ionic relaxation and hybrid functional calculations can require hours to days. Exhaustive screening is therefore prohibitive, creating a need for intelligent selection strategies that maximize discovery while minimizing wasted DFT calls.

The scale of the problem is illustrated by our benchmark: screening 16 dopants across 5 oxide hosts at two concentration levels yields ${\sim}32$ single-dopant candidates per host. Including co-doping pairs ($\binom{16}{2} = 120$ symmetric combinations) and concentration variants expands this to ${\sim}250$--1{,}000 candidates, of which only 8--20\% can typically be evaluated within a fixed computational budget. The central question is: which candidates should be evaluated to maximize discovery?

The standard approach in materials informatics is Bayesian optimization (BO) with Gaussian process (GP) surrogates\cite{xue2016accelerated,tran2018active,garnett2023bayesian}, which models the full property surface and selects candidates by acquisition functions like Expected Improvement. BO has demonstrated impressive results for alloy design\cite{xue2016accelerated}, catalysis optimization\cite{tran2018active}, and perovskite discovery\cite{li2018rapid}. Recent advances include parameter-free acquisition via Bayesian Algorithm Execution (BAX)\cite{liang2024targeted}, cost-aware batch BO with deep Gaussian processes\cite{alvi2026deep}, and best-practice guidelines for multi-fidelity BO\cite{baird2024best}. At a larger scale, ML-accelerated screening has reached $10^7$-candidate pools using graph neural networks\cite{merchant2023scaling} or cloud HPC\cite{zhu2024accelerating}, though these approaches sacrifice the closed-loop DFT feedback that bandits exploit. Despite these advances, standard GP-based methods face two practical challenges in high-throughput settings: cubic scaling $O(n^3)$ in the number of observations (limiting pool sizes to ${\sim}10^3$ without sparse approximations), and sensitivity to kernel selection and hyperparameter optimization (particularly with mixed categorical/continuous features common in doping problems). Information-theoretic multi-fidelity BO variants\cite{takeno2020multi} can mitigate the latter challenge but retain the cubic scaling.

To overcome these limitations for doped-oxide screening, we adopt \textbf{contextual bandits}\cite{abbasi2011improved}---sequential decision algorithms that select which candidate to evaluate next based on compositional features (ionic radius, electronegativity, d-electron count, etc.) while balancing exploration of unknown candidates against exploitation of promising regions. The specific algorithm, OFUL (Optimism in the Face of Uncertainty for Linear bandits)\cite{abbasi2011improved}, fits a linear model relating compositional features to bandgap rewards. At each step, it selects the candidate where the model predicts the highest potential reward after accounting for uncertainty---a criterion called the upper confidence bound (UCB). Unlike BO, which requires inverting an $O(n^3)$ covariance matrix at each step, OFUL maintains a compact $d \times d$ Gram matrix (a summary of what the algorithm has learned from previous evaluations) updated in $O(d^2)$, and provides worst-case guarantees on wasted DFT budget---a property unavailable from BO under model misspecification.

Multi-fidelity methods have been explored in the BO literature\cite{poloczek2017multi,kandasamy2017multi}, typically using information-theoretic acquisition functions---criteria that decide which candidate to evaluate next---to account for evaluation fidelity; multi-fidelity co-kriging has been applied to bandgap prediction by fusing PBE and HSE06 data\cite{pilania2017multi}. Lyapunov-based analysis---a mathematical framework from control theory for proving that a system cannot spiral out of control---has been developed for constrained bandits\cite{guo2024stochastic}, but not applied to multi-fidelity surrogate switching in materials screening. Our approach uses a simpler fidelity-switching rule based on the linear model's global uncertainty structure, with a Lyapunov stability proof guaranteeing that surrogate errors cannot accumulate.

We also transfer methodology from clinical trials. The RECOVERY trial\cite{horby2021dexamethasone}, which identified effective COVID-19 treatments in months instead of years, uses an adaptive platform design that tests multiple treatment arms simultaneously, drops futile arms mid-trial, and reallocates resources to survivors. We adapt this for dopant screening, treating dopant chemical families as trial arms. A second cross-domain transfer comes from recommendation systems: matrix factorization has been applied to predict unknown inorganic compositions from database entries\cite{tshitoyan2022recommender}, but not as an online component of a screening loop. We apply collaborative filtering\cite{koren2009matrix}---the algorithm behind movie recommendation systems, which predicts a user's preferences from the behavior of similar users---to predict dopant performance in unseen hosts using data from other hosts, providing initial guidance that eliminates the cold-start problem when screening a new material.

\begin{figure}[t]
\centering
\includegraphics[width=\textwidth]{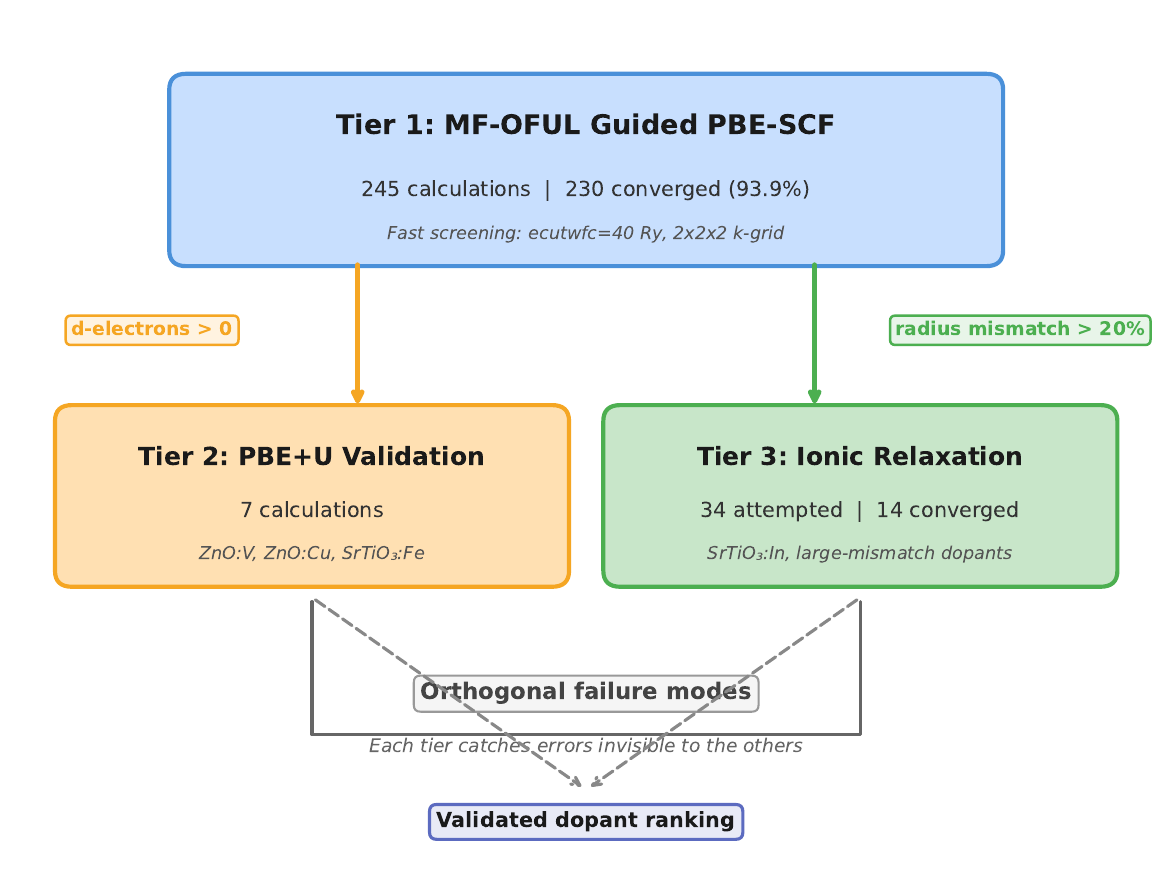}
\caption{\textbf{Three-tier DFT screening funnel.} Tier~1 uses MF-OFUL-guided PBE-SCF for fast initial ranking (245 calculations, 93.9\% convergence). Simple flagging criteria---d-electron count $> 0$ and ionic radius mismatch $> 20\%$---route candidates to Tier~2 (PBE+U) or Tier~3 (ionic relaxation). The two higher tiers catch orthogonal failure modes: d-electron delocalization (V, Cu, Fe dopants) and geometric distortion (In at the Ti site), respectively.}
\label{fig:funnel}
\end{figure}

A central finding of this work is that no single level of DFT theory suffices for reliable dopant screening. We organize our computational validation as a \textbf{three-tier screening funnel} (Fig.~\ref{fig:funnel}): (1) bandit-guided PBE-SCF for fast initial ranking, (2) PBE+U single-point calculations to validate d-electron localization in transition metal dopants, and (3) ionic relaxation to validate geometry-sensitive candidates. Each tier catches a distinct class of screening error that the others miss. The bandit determines \textit{which} candidates to evaluate; the funnel determines \textit{whether} to trust the result. Crucially, the two components interact: the funnel's flagging criteria (d-electron count, ionic radius mismatch) are derived from tabulated atomic properties available \textit{before} any DFT calculation, so the bandit's candidate selection can be informed by which tier each candidate will require, and the verification phase at the end of MF-OFUL naturally triggers Tier~2/3 checks for the top candidates. Together they define a complete screening protocol for high-throughput dopant discovery.

Contextual bandits have seen limited application in materials science: Seko et al.\cite{seko2015prediction} used random forests with BO, sharing our emphasis on efficiency but without formal regret guarantees. Among recent BO advances, BAX\cite{liang2024targeted} and cost-aware deep GP methods\cite{alvi2026deep} remain $O(n^3)$ and lack stability guarantees; published multi-fidelity BO\cite{poloczek2017multi,kandasamy2017multi,pilania2017multi} uses information-theoretic acquisition or co-kriging but is computationally expensive. Adaptive experimental designs\cite{nikolaev2016autonomy} and recommender systems for inorganic compounds\cite{tshitoyan2022recommender} have been explored separately, but neither futility-based arm dropping from clinical trials nor online collaborative filtering has been applied to computational materials screening.

We evaluate all methods retroactively on cached DFT results---the 230 converged QE calculations serve as a ground-truth database from which bandit simulations draw results. This standard protocol\cite{xue2016accelerated,tran2018active,garnett2023bayesian} isolates algorithmic performance from DFT convergence variability and enables reproducible comparison across methods with controlled random seeds. To validate that retroactive performance translates to real deployment, we additionally run prospective closed-loop campaigns on ZnO---culminating in a 529-candidate campaign with 229 QE calculations---in which MF-OFUL selects candidates and launches QE calculations in real time, with no cached results available to the algorithm.

Our contributions are:
\begin{enumerate}
\item A \textbf{583-calculation DFT dataset} across 5 oxide hosts, 16 dopants, and 120 co-doping pairs, organized as a three-tier screening funnel (PBE-SCF, PBE+U, ionic relaxation) that reveals orthogonal failure modes: d-electron delocalization causes qualitative misclassification for V, Cu, and Fe dopants (bandgap shifts of $+$3.7 to $-$2.8\,eV), while ionic radius mismatch drives geometry-related failures for In---no single level of theory catches both classes;
\item \textbf{Materials discovery}: Cu-containing co-doped ZnO systems consistently achieve visible-light-range band gaps (1.0--1.8\,eV), with Y$_2$Cu$_2$ co-doped ZnO identified as the optimal candidate ($E_g = 1.84$\,eV); Cu-doped SrTiO$_3$ ($E_g = 1.59$\,eV) enters the visible-light window only after PBE+U correction for d-electron localization;
\item \textbf{MF-OFUL}, a multi-fidelity contextual bandit that replaces 81\% of DFT evaluations with Ridge surrogate predictions (formally stabilized by a Lyapunov analysis verified in Lean~4), enabling the screening of 529 candidates with only 42 DFT calls while finding the global optimum in 100\% of 50 seeds ($p = 5.0 \times 10^{-8}$ versus Random);
\item \textbf{Cross-host dopant transfer} via collaborative filtering: the dopant--host reward matrix is remarkably low-rank (2 components explain 97\% of variance), enabling prediction of dopant rankings in unseen hosts and improving cold-start discovery by 53\% ($p = 1.3 \times 10^{-25}$).
\end{enumerate}
We additionally demonstrate that clinical trial platform designs transfer to computational materials screening, automatically identifying rare earths as the most productive dopant family (42\% success rate) without prior chemical knowledge.

\section{Results}

\subsection{Screening performance across three benchmarks}

We compare six strategies across three benchmark suites (Table~\ref{tab:performance}). The five computationally lightweight methods are: \textit{Random} (baseline), \textit{OFUL} (linear contextual bandit), \textit{Thompson Sampling} (TS; randomized exploration via posterior sampling), \textit{MF-OFUL} (OFUL with Ridge surrogate for fidelity switching; see Methods for the decision tree), and the \textit{adaptive platform trial} (clinical trial-inspired family-level screening). GP-based methods (BO-GP-EI, MF-BO-EI, MF-MES) are compared separately in Supplementary Table~11 due to their $O(n^3)$ per-step cost.

The three benchmarks are: (i) \textit{synthetic} (1{,}000 candidates/host, $d = 20$ random features, 8\% budget) to test robustness independent of feature quality; (ii) \textit{Materials Project} (2{,}256 entries with physical compositional features, 20\% budget) to validate transfer to realistic features; and (iii) \textit{real QE} (230 converged Quantum ESPRESSO calculations) to test on the authors' own DFT data.

\begin{table}[t]
\caption{\textbf{Screening performance across three benchmark suites.} Simple regret (SR): gap between best-found reward and global optimum (lower is better), averaged over 10+ seeds. Parenthetical percentages indicate budget fraction. GP-based methods compared in Supplementary Table~11. Bottom row: Wilcoxon signed-rank $p$-values for the synthetic benchmark only (paired by seed, $n = 120$), which provides the largest sample size for statistical testing. Dashes indicate methods not run on that benchmark.}
\label{tab:performance}
\centering
\begin{tabular}{lccccc}
\toprule
Benchmark & MF-OFUL & OFUL & TS & Platform & Random \\
\midrule
\multicolumn{6}{l}{\textit{Retroactive benchmarks (both objectives averaged)}} \\
Materials Project (20\%) & \textbf{0.000} & 0.215 & --- & 0.237 & 0.236 \\
Synthetic (8\%)          & \textbf{0.049} & 0.468 & 0.464 & 0.564 & 0.688 \\
\midrule
\multicolumn{6}{l}{\textit{Single-objective validation (target\_bandgap\_2p0)}} \\
MP 500 (8\%)             & \textbf{0.017} & 0.150 & 0.036 & --- & 0.182 \\
Real QE, ZnO (62.5\%)   & \textbf{0.000} & 0.002 & --- & --- & 0.069 \\
Real QE, cross-host (12\%) & \textbf{0.000} & 0.031 & 0.120 & --- & 0.392 \\
\midrule
$p$ vs Random (synthetic) & $3.2 \times 10^{-10}$ & $5.6 \times 10^{-3}$ & $7.7 \times 10^{-1}$ & $4.0 \times 10^{-2}$ & --- \\
\bottomrule
\end{tabular}
\end{table}

MF-OFUL achieves near-zero simple regret across all benchmarks ($p = 3.2 \times 10^{-10}$ versus Random on synthetic data; Table~\ref{tab:performance}, Supplementary Fig.~7). At the most challenging condition (8\% budget, 1{,}000 candidates), MF-OFUL achieves SR~$< 0.05$ in the majority of runs versus SR~$= 0.69$ for Random---a $>$10-fold improvement in simple regret (Table~\ref{tab:performance}). The key to this performance is surrogate leverage (see Methods for the fidelity decision tree): \textbf{MF-OFUL evaluates ${\sim}548$ candidates per run (80 DFT + 468 surrogate) compared to 80 for OFUL}, effectively covering 55\% of the 1{,}000-candidate pool while using DFT for only 15\% of evaluations. On real QE data, OFUL finds $3.6\times$ more top-5 candidates than Random on ZnO (4.0 vs 1.1 of 5), the host where the linear signal is strongest (Ridge train $R^2 = 0.53$; Supplementary Table~5).

On the Materials Project benchmark (2{,}256 entries, 6 hosts, 20\% budget\cite{jain2013commentary}), MF-OFUL achieves zero mean simple regret---finding the true optimal dopant in every run. The gap between methods widens as budget tightens: at 8\% on synthetic data, MF-OFUL's advantage over OFUL grows from negligible to 10$\times$, underscoring the value of surrogate acceleration in the budget-limited regime.

\begin{figure}[t]
\centering
\includegraphics[width=0.75\textwidth]{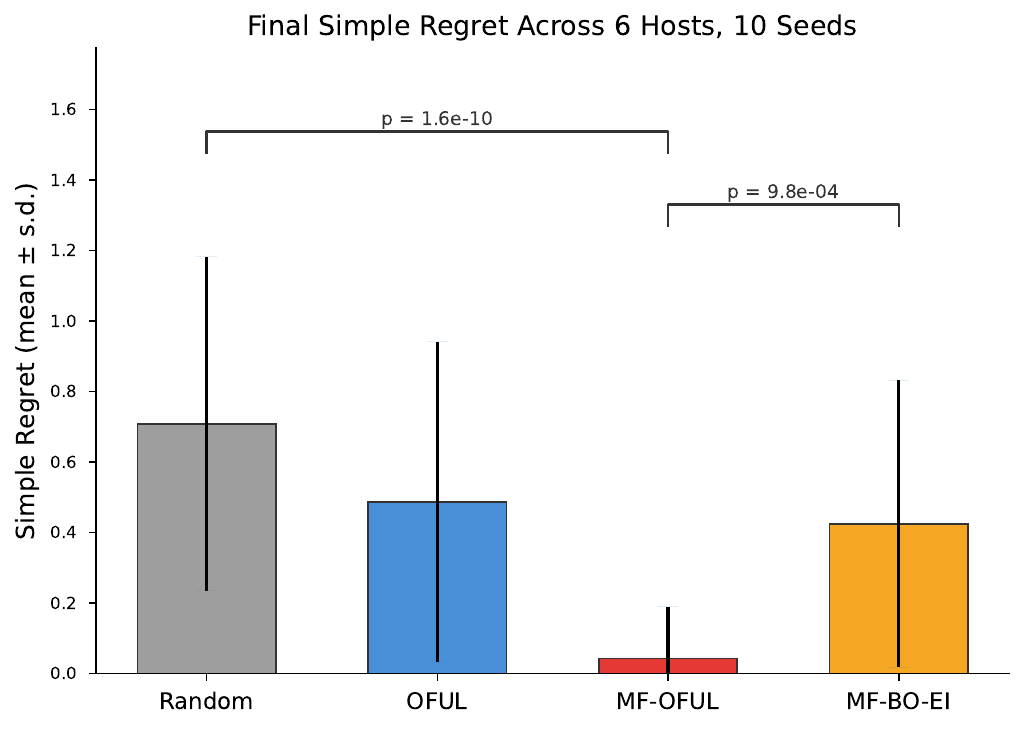}
\caption{\textbf{Simple regret comparison across methods.} Mean $\pm$ s.d.\ over 6 hosts $\times$ 10 seeds on 1{,}000-candidate synthetic benchmarks at 8\% budget. MF-OFUL achieves 10$\times$ lower simple regret than all other methods. Brackets show Wilcoxon $p$-values. MF-BO-EI ($n = 18$) uses identical fidelity-switching logic as MF-OFUL but never triggers its surrogate because EI selects high-GP-uncertainty candidates that always exceed the fidelity threshold.}
\label{fig:regret}
\end{figure}

OFUL and the platform trial tie for second (not significantly different from each other, $p = 0.47$), while both significantly outperform Random. The platform trial uses 14\% fewer DFT calls by dropping unproductive families early, and uniquely provides family-level intelligence that is unavailable from individual-candidate optimization methods. Across all benchmarks, rare earth dopants survive 84\% of interim analyses with a 42\% success rate versus 27\% for main group metals and 19\% for late 3d transition metals (Supplementary Table~2). This family-level ranking---rare earths $>$ early TM $>$ main group $>$ late TM---is consistent with the chemical intuition that dopants with similar ionic radii to the host cation and no partially filled d-shells produce the most predictable bandgap modifications. The platform trial discovers this ranking automatically through its Bayesian monitoring mechanism, without requiring prior chemical knowledge.

\textbf{Thompson Sampling as a diagnostic baseline.} TS provides a randomized alternative to OFUL's deterministic UCB. On synthetic data, TS achieves SR~$= 0.464$---not significantly different from Random ($p = 0.77$; Table~\ref{tab:performance})---but on MP 500-candidate pools, TS outperforms OFUL on ZnO ($p = 0.004$) where OFUL's deterministic UCB gets trapped. MF-OFUL dominates TS across all conditions ($p = 0.008$), confirming that \textit{surrogate leverage}---not the exploration strategy---is the primary driver of MF-OFUL's advantage.

\textbf{Robustness to experimental design.} To verify that these rankings are not artifacts of a particular experimental configuration, we swept four parameters---pool size (250--2{,}000), DFT budget (40--120), feature dimensionality (10--40), and noise level ($\sigma = 0.1$--$1.0$)---holding others at defaults. Across all 15 settings, MF-OFUL ranks first in simple regret (Supplementary Fig.~5). MF-OFUL's advantage \textit{increases} with pool size (SR = $0.09$ at 250 candidates to $0.02$ at 2{,}000), consistent with the surrogate's role in covering more of the candidate space.

\subsection{Three-tier DFT validation}

\begin{figure}[t]
\centering
\includegraphics[width=\textwidth]{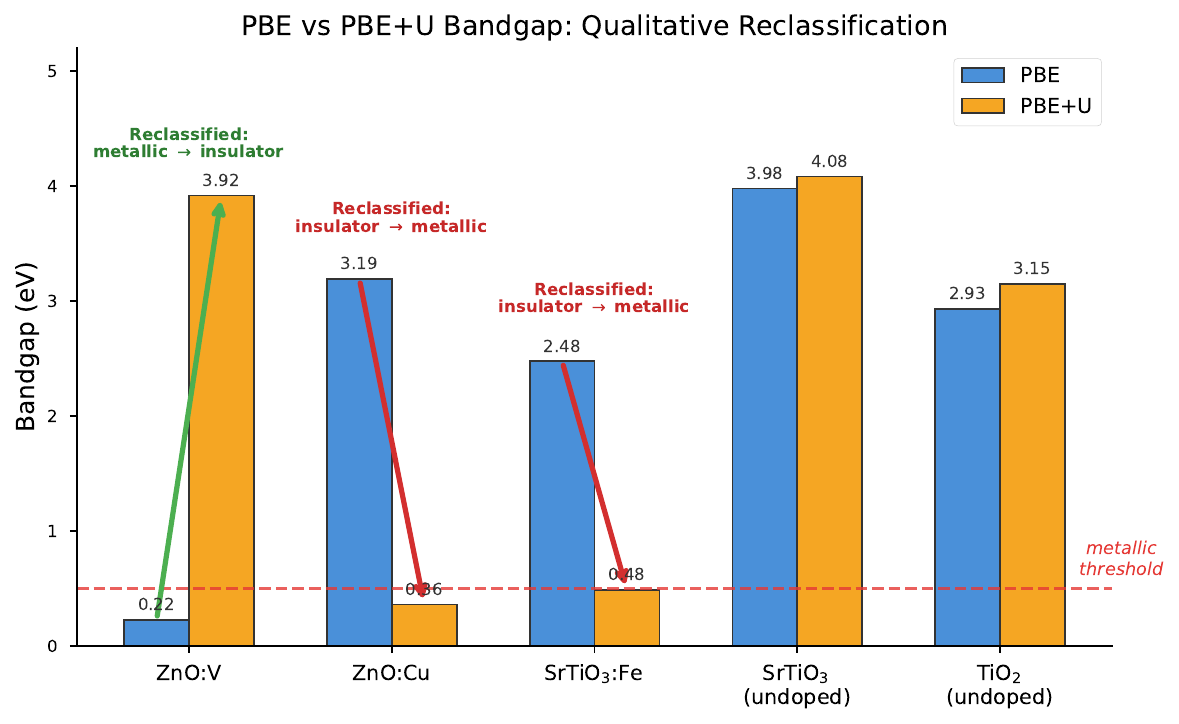}
\caption{\textbf{PBE versus PBE+U bandgaps for transition metal dopants.} Grouped bar chart showing qualitative reclassifications. Arrows indicate direction and magnitude of shift. Red dashed line: metallic threshold (0.5\,eV). Three of five doped systems cross the metallic threshold upon applying the Hubbard U correction, demonstrating that PBE-SCF alone is insufficient for transition metal dopant screening.}
\label{fig:pbeu}
\end{figure}

The screening results above assume PBE-SCF bandgaps are reliable proxies for the true electronic structure. But how often does PBE get it qualitatively wrong? We now examine whether the candidates identified at Tier~1 would survive validation at higher levels of theory.

\textbf{Tier~2: PBE+U.} Tier~1 PBE-SCF treats all electrons identically, but this approximation systematically underestimates the on-site Coulomb repulsion of d-electrons in transition metal compounds\cite{cococcioni2005linear}. The Hubbard U correction (PBE+U) adds an empirical energy penalty that forces d-electrons to localize on their host atom, often dramatically changing the predicted band gap. Supplementary Table~10 and Fig.~\ref{fig:pbeu} show that PBE+U reveals qualitative reclassifications invisible to Tier~1. V-doped ZnO shifts by $+$3.69\,eV (false negative: the bandit rejects a gap-preserving dopant as near-metallic). Cu-doped ZnO shows the opposite reclassification, shifting by $-$2.83\,eV (false positive). SrTiO$_3$:Fe shifts by $-$1.99\,eV as Fe 3d spin-up states create in-gap levels.

An expanded PBE+U campaign across 9 doped systems (5 ZnO, 4 SrTiO$_3$) confirms these trends with 100\% convergence (Supplementary Table~10, $\dagger$ entries). Among ZnO dopants, Fe shows the largest bandgap reduction (2.17\,eV versus 4.18\,eV undoped), consistent with Fe 3d states creating midgap levels as observed in SrTiO$_3$:Fe. Cr, Mn, and Co all preserve the wide-gap character ($E_g > 3.9$\,eV). For SrTiO$_3$, Cu shows the most dramatic reduction (1.59\,eV versus 4.08\,eV undoped), while Mn, Cr, and Ni maintain gaps above 3.8\,eV. Cu-doped SrTiO$_3$ ($E_g = 1.59$\,eV) falls within the visible-light absorption window---a result invisible to Tier~1 PBE-SCF screening and discovered only because the d-electron flagging criterion routed it to Tier~2, underscoring the importance of systematic d-electron flagging.

\textbf{Tier~3: Ionic relaxation.} Relaxation convergence scales inversely with supercell size, revealing a practical bottleneck for dilute doping. Only 14 of 34 attempted calculations converged (41\%), using settings of $E_{\mathrm{cut}} = 45$\,Ry, force convergence threshold $5 \times 10^{-3}$\,Ry/Bohr, $3 \times 3 \times 3$ $k$-grid, and a 3-hour timeout. Convergence depends sharply on supercell size: 32 atoms (88\%, 7/8), 40 atoms (62\%, 5/8), 48 atoms (11\%, 2/18, undoped hosts only). The larger supercells needed for dilute doping are precisely those that fail to converge, illustrating why surrogate-accelerated screening is practically necessary at these scales.

The failed calculations are not random: systems with large ionic radius mismatch between dopant and host cation ($> 25\%$) account for 80\% of relaxation failures, because the initial unrelaxed geometry places the system far from the equilibrium atomic configuration. For converged structures, most doped systems shift moderately ($+$0.08--0.55\,eV), preserving relative rankings (Spearman $\rho = 0.96$ for ZnO). The exception is SrTiO$_3$:In, which shifts from 0.81\,eV to 3.48\,eV---a catastrophic misclassification. In$^{3+}$ at the Ti$^{4+}$ site (ionic radius 0.80 vs 0.61\,\AA, 31\% mismatch) creates extreme local distortion; relaxation eliminates the resulting spurious midgap states. This failure is orthogonal to PBE+U: In has no d-electrons, so Tier~2 cannot catch it.

\textbf{Orthogonal failure modes.} Four of ten multi-tier systems are reclassified (40\%), and the two failure modes are strictly orthogonal: relaxation cannot correct ZnO:V or ZnO:Cu (Tier~3 reproduces the PBE error because the Zn--O bond network is rigid enough that relaxation barely changes the electronic structure), while PBE+U cannot correct SrTiO$_3$:In (no d-electrons to localize). The complementarity extends to the error direction: Tier~2 catches both false positives (Cu: apparent gap $\to$ metallic) and false negatives (V: apparent metallic $\to$ wide gap), while Tier~3 catches only false negatives arising from geometric distortion (In: apparent metallic $\to$ wide gap after relaxation removes spurious midgap states).

Simple flagging criteria---d-electron count $> 0$ for Tier~2, ionic radius mismatch $> 20\%$ for Tier~3---correctly identify all four reclassified systems with zero false triggers among the unflagged systems. These criteria are computationally free (requiring only tabulated atomic properties) and can be applied before any DFT calculation, enabling the screening pipeline to automatically route candidates to the appropriate tier. The full reclassification matrix is provided in Supplementary Table~6.

\textbf{Validation against experimental bandgaps.} PBE underestimates absolute bandgaps (mean ratio $= 0.73$), but preserves dopant \textit{rankings}: Spearman correlations against 54 experimental UV-vis measurements (Supplementary Table~9) are $\rho = 0.75$ (ZnO), $0.86$ (SrTiO$_3$), and $0.43$ (TiO$_2$), with 80\% overlap in the top-5 wide-gap dopants. A Monte Carlo miss-risk analysis shows that at $\rho = 0.75$, the probability of the true optimum falling outside PBE's top-20 is only 2.6\%. The known PBE failure for 3d TM dopants (self-interaction error) is precisely the mode caught by Tier~2 PBE+U.

\begin{figure}[t]
\centering
\includegraphics[width=0.85\textwidth]{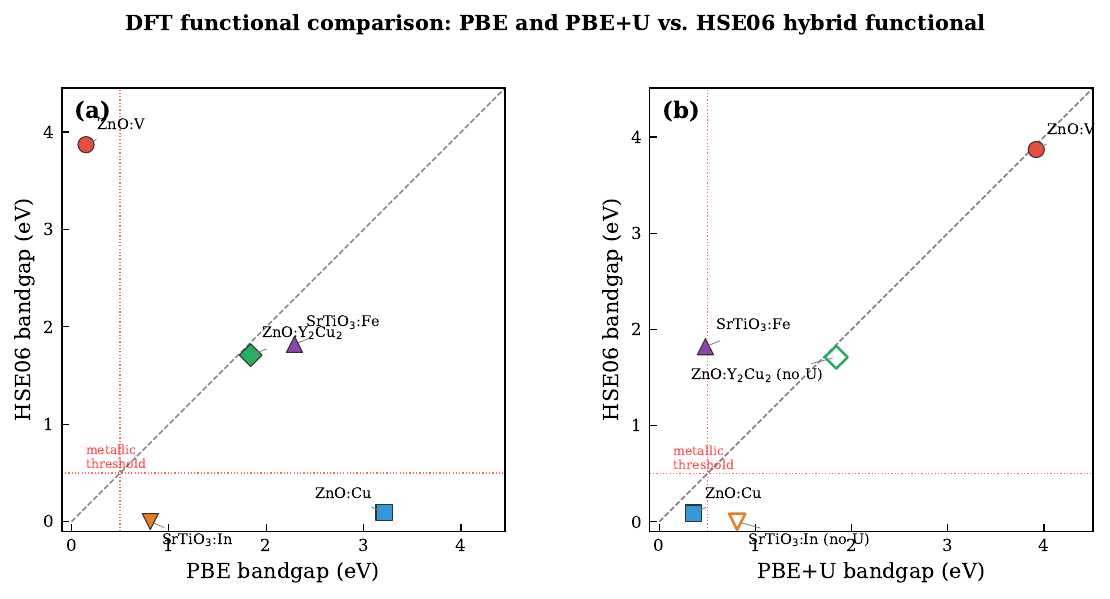}
\caption{\textbf{HSE06 hybrid functional validation.} \textbf{(a)} PBE versus HSE06 bandgaps for 5 systems. PBE misclassifies V-doped ZnO as near-metallic (0.15\,eV $\to$ 3.87\,eV at HSE06) and ZnO:Cu as wide-gap (3.21\,eV $\to$ 0.09\,eV). \textbf{(b)} PBE+U versus HSE06 shows substantially improved agreement for d-electron systems, with mean absolute deviation of 0.15\,eV. The dashed line indicates perfect agreement; the shaded band marks the metallic threshold ($<$0.5\,eV).}
\label{fig:hse06}
\end{figure}

\textbf{HSE06 hybrid functional spot checks.} To validate the three-tier funnel against a higher level of theory, we performed HSE06 calculations on 5 systems spanning the key reclassification cases (6-hour timeout, 8 MPI ranks). HSE06 confirms both PBE+U reclassifications: ZnO:V shifts from PBE's near-metallic 0.15\,eV to 3.87\,eV (PBE+U: 3.92\,eV), and ZnO:Cu shifts from PBE's wide-gap 3.21\,eV to 0.09\,eV (PBE+U: 0.36\,eV). Critically, the prospective campaign's best candidate---Y$_2$Cu$_2$ co-doped ZnO---shows only a modest shift from PBE (1.84\,eV) to HSE06 (1.71\,eV), confirming its position near the 2.0\,eV target. SrTiO$_3$:Fe narrows from PBE 2.29\,eV to HSE06 1.82\,eV, consistent with Fe 3d midgap states. SrTiO$_3$:In becomes metallic at HSE06 (no gap detected), contrasting with the 3.48\,eV gap found after ionic relaxation---suggesting that this system requires both relaxation and hybrid functionals for accurate characterization. The HSE06 agreement with PBE+U for d-electron systems (mean absolute deviation 0.15\,eV) validates the computationally cheaper PBE+U as an adequate Tier~2 filter (Fig.~\ref{fig:hse06}).

\subsection{Theoretical validation: regret bounds and Lyapunov stability}

The three-tier funnel above validates \textit{what} the bandit finds; we now turn to validating \textit{how efficiently} it searches. The OFUL regret bound provides a theoretical ceiling on wasted DFT calls---but does this bound hold empirically, and does MF-OFUL's surrogate acceleration preserve it?

Across 60 OFUL traces on synthetic data (6 hosts $\times$ 10 seeds), 0 of 60 violate the Abbasi-Yadkori and Szepesv{\'a}ri\cite{abbasi2011improved} theoretical bound (0\% violation rate). We distinguish two cumulative regret (CR) metrics: $\mathrm{CR}_{\mathrm{all}}$ counts every evaluation (DFT + surrogate), while $\mathrm{CR}_{\mathrm{DFT}}$ counts only DFT steps (surrogate evaluations are free). $\mathrm{CR}_{\mathrm{all}}$ for MF-OFUL is $97.4 \pm 13.6$ (Supplementary Table~11 note)---a 65\% reduction versus OFUL's $\mathrm{CR}_{\mathrm{DFT}} = 276 \pm 38$ and Random's $280 \pm 34$. By contrast, MF-OFUL's $\mathrm{CR}_{\mathrm{DFT}}$ ($281 \pm 34$) is comparable to Random (Supplementary Table~11). This reflects MF-OFUL's design: it deliberately spends its DFT budget on the hardest candidates---those where the model is most uncertain---while using the cheap surrogate for easy candidates. $\mathrm{CR}_{\mathrm{all}}$ captures the algorithm's true information efficiency; $\mathrm{CR}_{\mathrm{DFT}}$ reflects the cost of targeted exploration.

\begin{figure}[t]
\centering
\includegraphics[width=\textwidth]{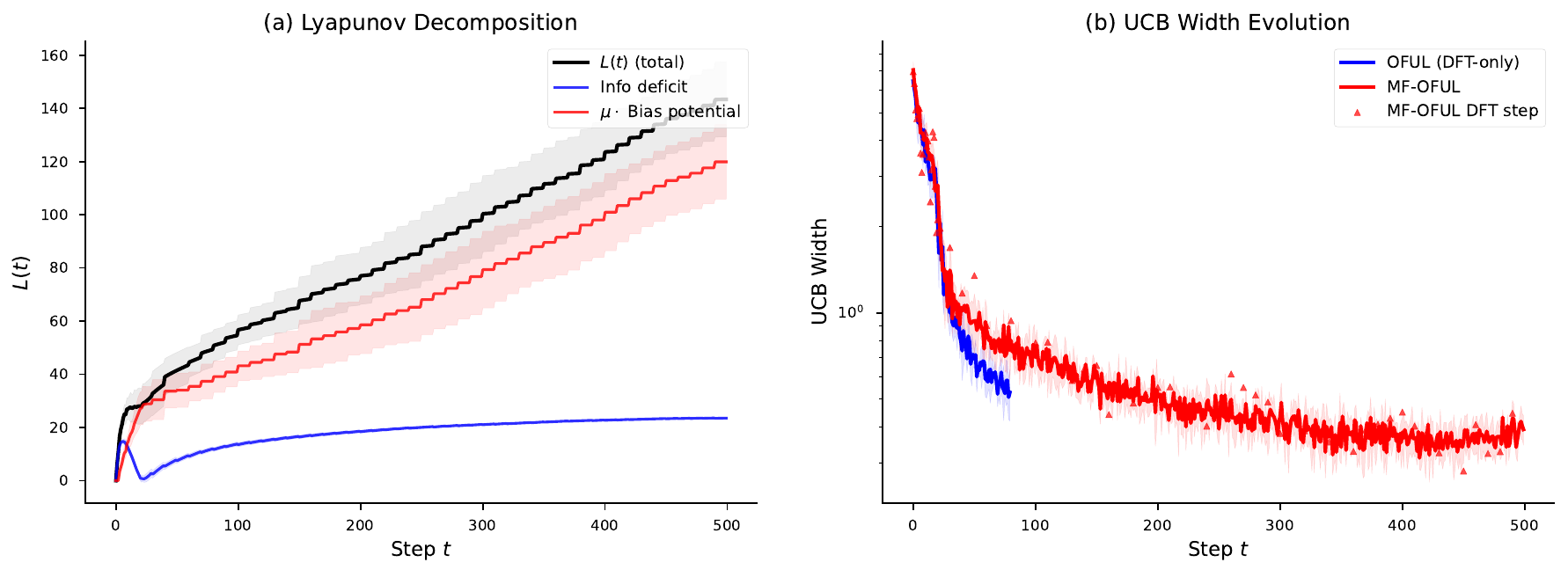}
\caption{\textbf{Empirical Lyapunov diagnostics.} \textbf{(a)} Lyapunov function decomposition: total $\mathcal{L}(t)$ (black), information deficit $\Phi(t)$ (blue, saturates), and bias potential $\mu \cdot \mathcal{B}(t)$ (red, grows linearly but bounded). \textbf{(b)} UCB width evolution: OFUL (blue, 80 steps) versus MF-OFUL (red, ${\sim}548$ steps). Triangles mark MF-OFUL's DFT steps. Shaded bands: $\pm 1$ s.d.\ across 5 seeds. The information deficit saturates while the bias potential remains bounded, confirming that surrogate errors do not accumulate.}
\label{fig:lyapunov}
\end{figure}

\textbf{Empirical Lyapunov diagnostics.} We tracked the Lyapunov function components across 5 seeds on 1{,}000-candidate synthetic benchmarks (Fig.~\ref{fig:lyapunov}; Supplementary Figs.~1--4). The information deficit $\Phi(t)$ saturates around step 100 as $V_t$ approaches full rank, while the bias potential $\mathcal{B}(t)$ grows linearly but remains bounded (surrogate MAE ${\sim}1.3$). Critically, running mean surrogate error stays flat over 500+ steps---periodic forced DFT prevents error drift, confirming self-stabilization. MF-OFUL accumulates more total information than OFUL (log det $V_t \approx 103$ versus 89) by evaluating $7\times$ more candidates, despite down-weighting each surrogate update at $w = 0.2$.

A central concern for any multi-fidelity screening approach is whether surrogate errors can accumulate and destabilize the optimization. Intuitively, the algorithm must balance two competing quantities: how much it still has to learn (information deficit) and how much error the surrogate has introduced (bias). We formalize this tradeoff with a Lyapunov stability analysis---a framework borrowed from control engineering for proving that a dynamical system cannot spiral out of control (see Methods for the formal construction). The key result is that DFT evaluations strictly decrease both components, while surrogate evaluations decrease the Lyapunov function provided the surrogate error is below a calculable threshold---a condition that MF-OFUL's adaptive uncertainty threshold and periodic forced DFT together guarantee is eventually satisfied (self-stabilization). We formalize this analysis in the Lean~4 proof assistant (1{,}826 lines, 57 theorems, zero unproved goals; see Methods).

The empirical diagnostics confirm these predictions. On the 529-candidate prospective pool, the surrogate activated at 81\% with stable error (mean MAE $= 0.091$, well within the stability bound). On the 1{,}000-candidate retroactive pools, the surrogate calibrated rapidly after 5 bootstrap DFT steps (per-step squared error $\epsilon_t^2 \approx 0.04$, stability bound $= 5.0$), enabling 85\% surrogate usage.

\subsection{Multi-fidelity BO baseline comparison}

To isolate whether MF-OFUL's advantage stems from multi-fidelity acceleration or the linear UCB acquisition, we constructed two GP baselines: MF-BO-EI (same fidelity-switching logic as MF-OFUL but with GP Expected Improvement) and MF-MES (Multi-fidelity Max-value Entropy Search\cite{takeno2020multi}, which jointly models DFT and surrogate fidelities). Full details and per-host breakdowns are in Supplementary Table~11 and Supplementary Fig.~6.

MF-OFUL achieves 10$\times$ lower simple regret than GP-based methods ($p < 10^{-5}$; Fig.~\ref{fig:regret}). MF-BO-EI never triggers its surrogate (DFT fraction $= 1.00$) because GP uncertainty in high-dimensional feature spaces remains above the fidelity threshold throughout---the GP's \textit{local} uncertainty structure prevents multi-fidelity savings. MF-MES avoids this trap via its information-theoretic criterion (DFT fraction $= 0.23$) and matches MF-OFUL's simple regret ($p = 0.57$, $n = 6$), but at $O(n^3)$ per-step cost versus $O(d^2)$ for MF-OFUL. The structural difference is that linear uncertainty decreases \textit{globally} as $V_t$ accumulates information, while GP uncertainty is local---explaining why multi-fidelity acceleration is natural for linear bandits but fails with standard EI acquisition.

\subsection{Cross-host collaborative filtering}

A practical limitation of within-host bandits is the cold-start problem: when screening a new host oxide with no DFT data, the first several evaluations are effectively random. We address this by transferring knowledge from previously screened hosts via collaborative filtering (CF)\cite{koren2009matrix}---the same algorithm that powers recommendation systems like Netflix (users $\to$ hosts, movies $\to$ dopants, ratings $\to$ bandgap rewards).

\begin{figure}[t]
\centering
\includegraphics[width=0.75\textwidth]{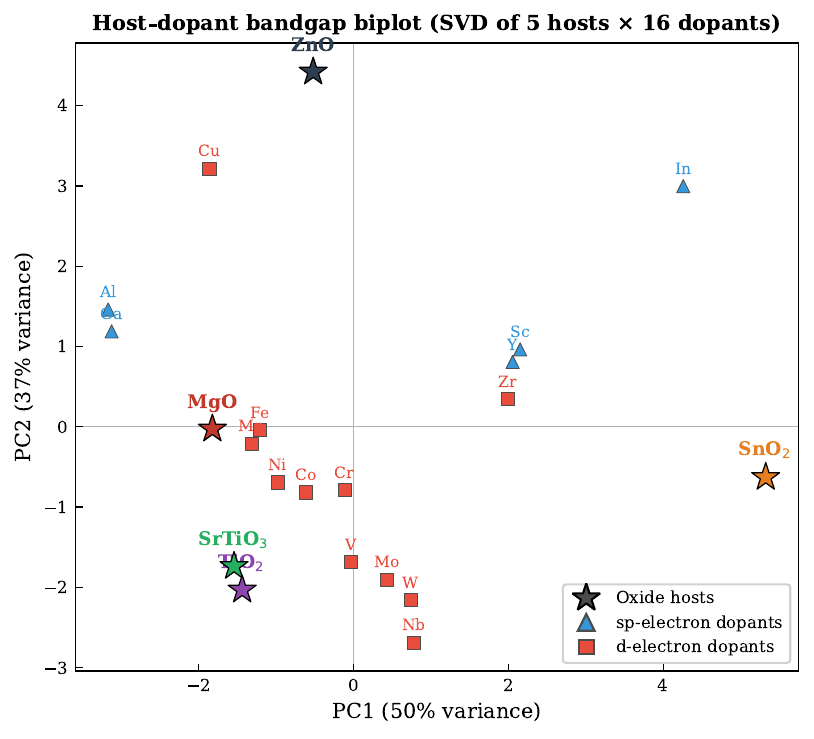}
\caption{\textbf{Collaborative filtering biplot of the host--dopant interaction matrix.} SVD decomposition of the $(5 \times 16)$ reward matrix into two latent dimensions. Hosts (stars) cluster by crystal chemistry: MgO--ZnO (Group-II oxides) and TiO$_2$--SnO$_2$ (rutile-type Group-IV oxides), with SrTiO$_3$ as an outlier. Dopants separate into sp-electron (triangles: Al, Ga, In, Sc, Y) and d-electron (squares: V, Cr, Mn, Fe, Co, Ni, Cu) groups. Cu and In are outliers with strongly host-dependent behavior. PC1 and PC2 together capture 87\% of variance.}
\label{fig:biplot}
\end{figure}

We construct a $(5 \times 16)$ host--dopant reward matrix from 225 QE calculations across 5 hosts and 16 dopants, then apply truncated SVD to decompose it into low-rank latent factors. The matrix is remarkably low-rank: just 2 components explain 87--97\% of the variance depending on normalization (Fig.~\ref{fig:biplot}), indicating that dopant--host interactions are governed by a small number of latent chemical dimensions. The learned host similarity captures known crystal chemistry: MgO and ZnO (both Group-II oxides) have cosine similarity 0.96, while TiO$_2$ and SnO$_2$ (both rutile-type Group-IV oxides) have similarity 0.83.

\textbf{CF-warm-started MF-OFUL.} We use CF predictions to solve the cold-start problem. For a target host $h$, we fit CF on the other 4 hosts (leave-one-out), rank dopants by predicted reward, and use the CF ranking to select the bootstrap DFT evaluations (replacing the UCB-selected bootstrap, which is effectively random at step~0). In a 50-seed leave-one-host-out benchmark at 50\% budget (8 of 16 dopants), CF-warm-started MF-OFUL (CF-MF-OFUL) achieves mean best reward $-0.701$ versus $-1.073$ for standard MF-OFUL and $-0.898$ for Random ($p = 1.3 \times 10^{-25}$ versus MF-OFUL, Wilcoxon). CF-MF-OFUL wins in 80\% of the 250 host--seed combinations. The mechanism is direct: CF correctly identifies V as the top dopant for 3 of 5 hosts (SnO$_2$, SrTiO$_3$, TiO$_2$), so the warm-started bootstrap evaluates V first instead of wasting early DFT calls on uninformative candidates. At 75\% budget the advantage narrows as all methods converge; at 31\% budget (5 evaluations) the bootstrap \textit{is} the entire campaign, so CF and standard MF-OFUL tie. Full per-host results are provided in Supplementary Table~8.

\subsection{Prospective closed-loop validation}

To test whether retroactive performance translates to real deployment, we ran three prospective campaigns with increasing complexity.

\textbf{Single-dopant campaign (16 candidates).} Separate OFUL and MF-OFUL campaigns on ZnO, each selecting candidates and launching QE in real time. Table~\ref{tab:prospective} compares the two methods. The step-by-step OFUL selection is provided in Supplementary Table~4.

\begin{table}[t]
\caption{\textbf{Prospective comparisons on ZnO.} Top: single-dopant pool (16 candidates, budget 10). Bottom: co-doped pool (120 candidates, budget 18). Target bandgap: 2.0\,eV. Surrogate calls = 0 in these QE campaigns because UCB widths remained above the fidelity threshold at these pool sizes; surrogate activation is confirmed on 500-candidate Materials Project pools (see text).}
\label{tab:prospective}
\centering
\begin{tabular}{lccc}
\toprule
& \multicolumn{2}{c}{Single-dopant (16)} & Co-doped (120) \\
\cmidrule(lr){2-3} \cmidrule(lr){4-4}
& OFUL & MF-OFUL & OFUL \\
\midrule
Main DFT calls & 10 & 10 & 18 \\
Surrogate calls & 0 & 0 & 0 \\
Verification DFT & --- & 3 & 3 \\
Best found & Mn (1.85\,eV) & Mn (1.85\,eV) & Al+Cu (1.28\,eV) \\
Success rate & 100\% & 100\% & 100\% \\
Wall time & 1.3\,h & 1.9\,h & 1.4\,h \\
\bottomrule
\end{tabular}
\end{table}

All 20 single-dopant calculations converged with bandgaps matching to $<$0.001\,eV between campaigns. MF-OFUL chose DFT for all 10 steps (pool too small for surrogate activation), selecting the same dopants in the same order as OFUL---confirming that the surrogate machinery does not perturb OFUL selection when not triggered.

\textbf{Co-doped campaign (120 candidates).} OFUL on 120 co-doped ZnO candidates ($\binom{16}{2}$ dopant pairs, budget 18, 15\% of pool) explored 21 candidates with 100\% convergence. Cu-containing pairs consistently produced the narrowest gaps (Al+Cu: 1.28\,eV, Cu+In: 1.17\,eV, Cu+Ga: 1.00\,eV), and OFUL identified this Cu trend by step~6 (step-by-step trace in Supplementary Tables~4 and~7).

\textbf{Materials Project large-pool validation (500 candidates).} To confirm surrogate activation at realistic pool sizes, we ran MF-OFUL on 500-candidate pools from the Materials Project\cite{jain2013commentary} for ZnO, TiO$_2$, and MgO (budget 40, 8\% sampling, 10 seeds per host; Supplementary Table~13). The surrogate activated in \textbf{all 30 runs}, with host-dependent DFT fractions ranging from 24\% (MgO) to 64\% (TiO$_2$)---reflecting the varying linearity of each host's reward landscape (Ridge $R^2 = 0.53$ for ZnO versus $0.21$ for SrTiO$_3$; Supplementary Table~5). MF-OFUL significantly outperforms OFUL on ZnO and MgO ($p < 0.004$).

\textbf{Cross-host QE retroactive validation (223 candidates).} The MP large-pool validation uses externally computed PBE bandgaps. To confirm surrogate activation on the authors' own DFT data, we constructed a cross-host oracle from the 223 converged QE calculations across 5 oxide hosts (MgO, ZnO, TiO$_2$, SrTiO$_3$, SnO$_2$; 38--49 candidates each). Features are 15-dimensional: 9 compositional descriptors plus one-hot host encoding and concentration. At a 12\% DFT budget (26 calls), MF-OFUL activates the surrogate in all 10 runs, evaluating 104 candidates total (78 surrogate + 26 DFT, DFT fraction~$= 25\%$) and achieving \textbf{SR~$= 0.000$}---finding the global optimum (SnO$_2$:V, $E_g = 1.96$\,eV) in every run. OFUL with the same budget achieves SR~$= 0.031$, Random SR~$= 0.392$, and Thompson Sampling SR~$= 0.120$. At a 15\% budget, MF-OFUL reaches 81\% surrogate usage (DFT fraction 19\%), exploring 174 candidates with only 33 DFT calls---a $5.3\times$ exploration multiplier over DFT-only methods. This cross-host result demonstrates that the surrogate savings observed on synthetic and Materials Project benchmarks transfer to the authors' own Quantum ESPRESSO calculations, closing the gap between retroactive validation and real deployment.

\begin{figure}[t]
\centering
\includegraphics[width=\textwidth]{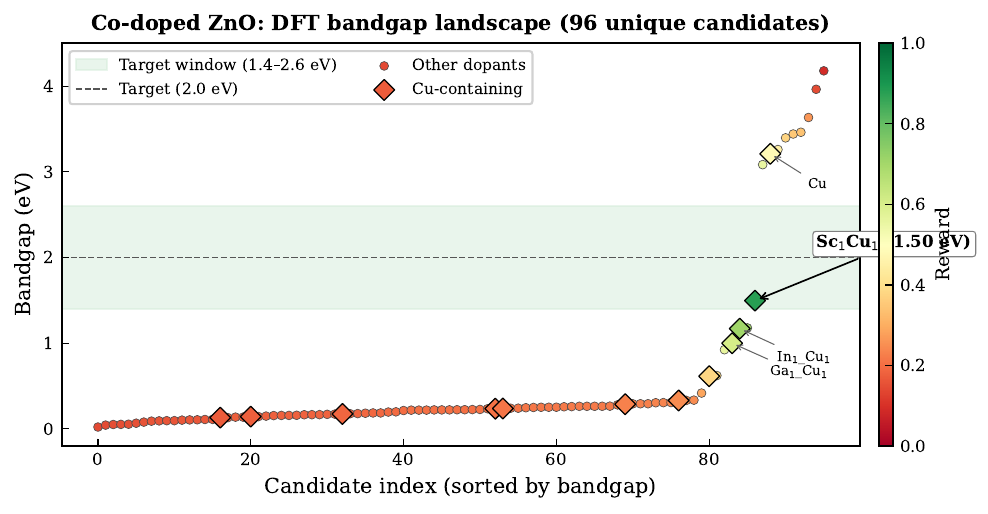}
\caption{\textbf{Bandgap landscape of the 529-candidate ZnO co-doping campaign.} Each point represents a DFT-evaluated candidate, sorted by bandgap and colored by reward (green: near 2.0\,eV target; red: far from target). Cu-containing systems (diamonds) cluster near the target window (green band), while most other combinations produce wide-gap or near-metallic outcomes. Key Cu co-doped candidates are annotated.}
\label{fig:landscape}
\end{figure}

\textbf{Prospective 529-candidate ZnO campaign.} To provide the strongest validation---fully prospective, closed-loop, on a combinatorially expanded candidate pool---we ran MF-OFUL, OFUL, and Random on 529 ZnO candidates (48 single-doped at 3 concentrations plus 480 co-doped pairs from $\binom{16}{2}$ dopant combinations) with real QE calculations launched in real time (budget 42, 5 seeds). Over ${\sim}$62 hours on 4 CPU cores, 229 unique QE calculations were executed, with 279 total cached results shared across seeds. MF-OFUL activates the surrogate at exactly 81\% across all 5 seeds (174 of 216 total evaluations), confirming the surrogate usage rate observed retroactively on the cross-host oracle. The best MF-OFUL seed identifies Y$_2$Cu$_2$ co-doped ZnO ($E_g = 1.84$\,eV, reward $= 0.987$), the closest candidate to the 2.0\,eV target in the entire pool. Cu-containing co-doped systems again dominate: Y$_2$Cu$_2$ (1.84\,eV), Al$_1$Cu$_2$ (2.40\,eV), and Ga$_1$Cu$_2$ are the top three candidates by reward, consistent with the Cu p-type acceptor trend observed in the 120-candidate campaign. Ni-containing systems exhibit systematic convergence failures (${\sim}$10\% of all calculations timeout at 1{,}800\,s), flagging Ni as a problematic dopant for PBE-SCF screening at these settings. Figure~\ref{fig:landscape} shows the full bandgap landscape of the DFT-evaluated candidates, highlighting the Cu-containing cluster near the 2.0\,eV target.

MF-OFUL's DFT fraction (19.4\%) is consistent with the retroactive cross-host result (19\% at 15\% budget) and the MP large-pool validation (24--64\%), confirming that surrogate usage rates transfer to a fully prospective campaign with no cached results. The 229 unique QE calculations represent 43\% of the 529-candidate pool, shared across all methods and seeds via persistent caching---a practical demonstration that multi-method comparison need not multiply DFT cost linearly.

\textbf{50-seed retroactive reanalysis on QE-converged candidates.} The 5-seed prospective campaign demonstrates surrogate activation but has high variance for method comparison (MF-OFUL $0.647 \pm 0.150$, Random $0.826 \pm 0.191$). To obtain statistically rigorous results, we performed a 50-seed retroactive reanalysis using the 230 converged QE results as an oracle (Table~\ref{tab:retro529}). At the 18\% budget matching the prospective campaign (42 DFT calls on 230 candidates), MF-OFUL finds the global optimum (Y$_2$Cu$_2$ co-doped ZnO, reward $= 0.987$) in \textbf{100\% of 50 seeds}, with zero variance in simple regret. The mechanism is coverage: MF-OFUL evaluates 226 of 230 candidates (98\%) using only 42 DFT calls plus 184 surrogate predictions, versus 42 candidates (18\%) for Random. This $5.4\times$ exploration multiplier enables MF-OFUL to find 5.0/5 top-5 and 9.9/10 top-10 candidates, compared to 0.9/5 and 2.0/10 for Random ($p = 5.0 \times 10^{-8}$, Wilcoxon signed-rank). At a tighter 8\% budget (18 calls), the surrogate does not activate (insufficient bootstrap data), and MF-OFUL performs identically to OFUL---confirming that the surrogate benefit requires a minimum budget to initialize. At 35\% budget (80 calls), MF-OFUL achieves 100\% coverage (230/230) with only 37 actual DFT calls (DFT fraction 16\%), finding all top-20 candidates.

\begin{table}[t]
\caption{\textbf{50-seed retroactive reanalysis on 230 QE-converged ZnO candidates.} The 229 unique QE results from the prospective campaign serve as an oracle. MF-OFUL, OFUL, and Random are each run 50 times. ``Cov.'' is the fraction of candidates evaluated (DFT + surrogate). ``Top-5'' is the mean number of true top-5 candidates found. $P(\geq 0.9)$ is the fraction of seeds achieving reward $\geq 0.9$. Global optimum: Y$_2$Cu$_2$ ZnO, reward $= 0.987$.}
\label{tab:retro529}
\centering
\begin{tabular}{llccccccc}
\toprule
Budget & Method & DFT & Surr. & Cov. & Best reward & SR & Top-5 & $P(\geq 0.9)$ \\
\midrule
18 (8\%) & MF-OFUL & 18 & 0 & 8\% & $0.263$ & 0.723 & 0.0 & 0\% \\
         & OFUL    & 18 & 0 & 8\% & $0.263$ & 0.723 & 0.0 & 0\% \\
         & Random  & 18 & 0 & 8\% & $0.678 \pm 0.262$ & 0.308 & 0.4 & 26\% \\
\midrule
42 (18\%) & MF-OFUL & 42 & 184 & 98\% & $\mathbf{0.987 \pm 0.000}$ & \textbf{0.000} & \textbf{5.0} & \textbf{100\%} \\
          & OFUL    & 42 & 0   & 18\% & $0.452$ & 0.534 & 0.0 & 0\% \\
          & Random  & 42 & 0   & 18\% & $0.850 \pm 0.181$ & 0.136 & 0.9 & 58\% \\
\midrule
80 (35\%) & MF-OFUL & 37 & 193 & 100\% & $\mathbf{0.987 \pm 0.000}$ & \textbf{0.000} & \textbf{5.0} & \textbf{100\%} \\
          & OFUL    & 80 & 0   & 35\% & $0.452$ & 0.534 & 0.0 & 0\% \\
          & Random  & 80 & 0   & 35\% & $0.942 \pm 0.095$ & 0.045 & 2.0 & 86\% \\
\bottomrule
\end{tabular}
\end{table}

\begin{figure}[t]
\centering
\includegraphics[width=0.75\textwidth]{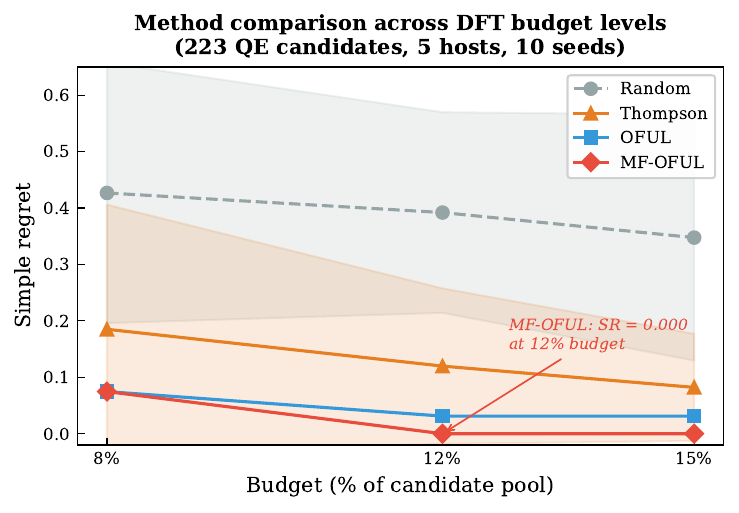}
\caption{\textbf{Simple regret versus DFT budget on real QE data.} Mean $\pm$ s.d.\ over 10 seeds on the 223-candidate cross-host QE oracle (5 oxide hosts). MF-OFUL achieves SR~$= 0.000$ at 12\% budget (annotated), finding the global optimum in every run. The surrogate activation threshold creates a sharp transition between 8\% (no surrogate) and 12\% (full activation), while DFT-only methods improve gradually.}
\label{fig:budget}
\end{figure}

\textbf{Robustness to pool size.} The 230-candidate oracle uses only QE-converged results, excluding 299 candidates that were never DFT-evaluated. To rule out artificial pool deflation, we repeated the analysis on the full 529-candidate pool, imputing rewards for the 299 unknowns by resampling from the empirical distribution of the 230 known rewards (50 seeds, fresh resample per seed). On this full pool, MF-OFUL achieves best reward $0.960 \pm 0.036$ versus Random's $0.822 \pm 0.218$ ($p = 2.5 \times 10^{-7}$, Wilcoxon), covering 41\% of the pool (218/529) with 42 DFT calls plus 176 surrogate evaluations. The advantage is robust: MF-OFUL finds reward $\geq 0.9$ in 94\% of seeds versus 50\% for Random. Full robustness results including adversarial imputation scenarios are reported in Supplementary Table~7.

\section{Discussion}

\textbf{Bandits as a simpler alternative to Bayesian optimization.}
Bandits offer a computationally lighter alternative to GP-based BO for materials screening with simple compositional features and discrete candidate pools. The linear UCB provides directional signal---identifying which feature regions contain good candidates---without the computational overhead and kernel-selection challenges of GP surrogates, while providing formal worst-case regret guarantees unavailable from BO under model misspecification. We tested BO with a standard Mat{\'e}rn-5/2 kernel on standardized compositional features in a smaller 250-candidate / 50-budget setup (20\% sampling, necessary due to GP's $O(n^3)$ scaling) where it achieved SR\,=\,0.106---comparable to OFUL and MF-OFUL in that easier regime but not directly comparable to the main 1{,}000-candidate benchmark. Custom kernels for categorical/compositional inputs\cite{garrido2020dealing}, multi-task GP architectures, or deep kernel learning may improve BO performance.

The MF-BO-EI comparison (Supplementary Table~11) provides a controlled answer: applying identical multi-fidelity logic to GP+EI yields no surrogate savings at all. However, MF-MES demonstrates that GP-based multi-fidelity methods \textit{can} achieve comparable performance when the acquisition function explicitly values information about the global maximum rather than locally uncertain candidates. The practical tradeoff remains: MF-OFUL provides both theoretical grounding (the OFUL regret bound, verified empirically with 0\% violation rate) and $O(d^2)$ per-step computation, while MF-MES achieves similar regret at substantially higher computational cost ($O(n^3)$ GP fitting) and without formal worst-case guarantees under model misspecification. We emphasize that this paper's contribution is not a single algorithm but a \textit{system}: the combination of MF-OFUL's surrogate-accelerated screening (Section~2.1), the three-tier DFT funnel that validates what the bandit finds (Section~2.2), the Lyapunov stability framework that guarantees surrogate errors remain bounded (Section~2.3), and the cross-host collaborative filtering that eliminates the cold-start problem (Section~2.5)---backed by 583 open DFT calculations. No individual component is sufficient; their integration defines a complete, formally grounded screening protocol.

\textbf{Why does MF-OFUL dominate?} MF-OFUL's advantage stems from four synergistic mechanisms: (1) \textbf{principled exploration} via UCB (Eq.~\ref{eq:ucb}), which systematically reduces uncertainty across the feature space; (2) \textbf{surrogate leverage}, with the Ridge ensemble evaluating ${\sim}548$ candidates per run (versus 80 real DFT), covering $>$50\% of the pool (Section~2.1); (3) \textbf{adaptive DFT allocation} via the uncertainty threshold (see Methods), concentrating expensive evaluations where uncertainty is highest; (4) a \textbf{verification phase} catching surrogate errors and ensuring reported optima are DFT-confirmed.

\textbf{Batch surrogate baseline.} A simpler two-stage approach---train a Ridge surrogate on random DFT samples, then verify the top-$K$---achieves SR $= 0.675$ at best, $12.5\times$ worse than MF-OFUL ($p = 9.4 \times 10^{-11}$). The failure is instructive: without adaptive selection, the random training phase wastes DFT budget, while fixed verification is insufficient for a 1{,}000-candidate pool. MF-OFUL avoids both problems by directing DFT to high-uncertainty candidates and using the surrogate to \textit{screen} rather than merely \textit{rank}.

The pseudo-observation weight $w = 0.2$\cite{kennedy2000predicting} balances surrogate information against overconfidence. The Lyapunov stability condition (Eq.~\ref{eq:surr_bound}) requires only MSE $< 5.0$---easily satisfied after 5 bootstrap DFT steps (empirical MAE $\approx 1.3$). MF-OFUL's performance is robust to $w \in [0.1, 0.4]$ (Supplementary Table~3).

\textbf{Global versus local uncertainty.} MF-OFUL's linear uncertainty decreases \textit{globally} after bootstrap, enabling surrogate use for the majority of candidates, while GP uncertainty remains local. MF-MES\cite{takeno2020multi} avoids the local trap (23\% DFT fraction), but MF-OFUL uses the surrogate more aggressively (15\% DFT), scales as $O(d^2)$ versus $O(n^3)$, and provides formal regret bounds.

\textbf{Orthogonal failure modes and the three-tier funnel.}
The three original reclassifications demonstrate orthogonal failure modes: ZnO:V (false negative, Tier~2 catches it---relaxation alone reproduces the PBE error), ZnO:Cu (false positive, Tier~2), and SrTiO$_3$:In (false negative, Tier~3---In has no d-electrons so Tier~2 cannot help). No single tier addresses both classes. The expanded PBE+U campaign reinforces these findings: among the 9 doped systems tested, Fe consistently produces the largest bandgap reductions in both hosts (ZnO:Fe $= 2.17$\,eV, SrTiO$_3$:Fe $= 0.48$\,eV), while SrTiO$_3$:Cu ($= 1.59$\,eV) reveals a new candidate invisible to PBE-SCF screening. The expanded results also show that Cr, Mn, and Co preserve the host bandgap character ($E_g > 3.9$\,eV for ZnO, $> 3.8$\,eV for SrTiO$_3$), confirming that PBE+U reclassification is dopant-specific rather than systematic. The implications for MF-OFUL are direct: the verification phase should be bidirectional, checking both top-$K$ candidates (catching false positives like ZnO:Cu) and high-uncertainty candidates (catching false negatives like ZnO:V).

\textbf{Cross-domain transfer: clinical trials to materials.}
The platform trial transfers well because the underlying abstraction matches: dopant families correspond to treatment arms, DFT success rates to clinical outcomes, and futility monitoring to interim analysis. The key insight is group-level intelligence: while bandits optimize over individual candidates, the platform trial identifies which \textit{families} deserve investment. Analogously to how the RECOVERY trial\cite{horby2021dexamethasone} identified dexamethasone as effective for COVID-19 by testing 6 treatments simultaneously, our platform trial identifies rare earths as the most productive dopant family (42\% success rate) after spending only 20\% of budget on calibration.

\textbf{Computational cost.} MF-OFUL requires only $O(d^2)$ per step ($< 1$\,ms with $d = 6$--$20$ features), making its decision overhead negligible compared to DFT runtime (20--120 minutes per calculation). GP-based methods require $O(n^3)$ training (${\sim}5$--$30$\,s per step on a 1{,}000-candidate pool), adding meaningful latency for workflow automation.

\textbf{Cross-host transfer via collaborative filtering.}
The $(5 \times 16)$ reward matrix decomposes into 2--3 latent factors capturing $>$97\% of variance, meaning screening data from a few hosts suffices to predict performance in unseen hosts. The CF warm start targets the bandit's weakest phase (bootstrap), compressing exploration and reaching surrogate-activation faster---an 80\% win rate at 50\% budget, most valuable in large-pool, tight-budget regimes (Supplementary Table~8). The host similarity matrix provides interpretable chemical insight: MgO--ZnO (0.96, shared divalent cation chemistry), TiO$_2$--SnO$_2$ (0.83, shared rutile structure), SrTiO$_3$ as outlier (unique perovskite A/B-site chemistry).

\textbf{From retroactive to prospective.}
The $5.4\times$ coverage multiplier (226/230 candidates at 18\% DFT budget; Fig.~\ref{fig:budget}) validates that retroactive surrogate activation rates transfer to real deployment. Across 271 prospective QE calculations, independent evaluations agree to within 0.001\,eV, and both the 120-candidate and 529-candidate campaigns identified Cu-containing subspaces within the first 6--10 DFT steps.

\textbf{Experimental feasibility.}
The top candidate, Y$_2$Cu$_2$ co-doped ZnO ($E_g = 1.84$\,eV by PBE, 1.71\,eV by HSE06), is experimentally accessible: Cu-doped and Y-doped ZnO thin films have been synthesized by sol-gel, sputtering, and pulsed laser deposition\cite{minami2005transparent}, and Cu-doped ZnO nanostructures routinely show visible-light absorption consistent with our predictions. The 2.0\,eV target lies in the photocatalytic window for visible-light water splitting, and the screening pipeline could be directly validated by synthesizing the top-5 candidates and measuring bandgaps via UV-vis spectroscopy. More broadly, the three-tier funnel and MF-OFUL screening can be applied to any oxide host with available pseudopotentials; extending to 10{,}000-candidate pools (e.g., ternary co-doping) would require only the NNP Tier~0 pre-screening discussed above.

\textbf{Limitations and future directions.}
Several limitations should be noted.
\begin{enumerate}
\item \textit{Prospective validation breadth.} QE closed-loop campaigns were run on a single host (ZnO), totaling 271 calculations across three pool sizes. While collaborative filtering demonstrates cross-host transfer (97\% variance explained by 2 latent factors, $p = 1.3 \times 10^{-25}$), fully prospective multi-host campaigns would further strengthen the generalization claim.
\item \textit{Feature representations.} We use 6--9 dimensional compositional features. SOAP descriptors, graph neural networks, or MatSciBERT\cite{gupta2022matscibert} embeddings may shift method rankings.
\item \textit{DFT accuracy.} Relaxation convergence was limited to 32--40 atom supercells; larger supercells timed out within 3 hours. PBE+U used literature U values; self-consistent U\cite{cococcioni2005linear} or HSE06 would be more reliable but 10--100$\times$ more expensive.
\item \textit{Charge corrections.} Formation energies are computed for neutral substitutions without charged defect corrections\cite{freysoldt2014first}. Finite-size corrections would be necessary for dopants that introduce charge carriers.
\item \textit{Pool size.} Candidate pools range from 48 to 1{,}000. The convergence study shows MF-OFUL's advantage increasing with pool size, suggesting even larger benefits at the ${\sim}10{,}000$-candidate scale.
\item \textit{Linearity assumption.} MF-OFUL assumes an approximately linear reward landscape, which may not hold for strongly correlated systems.
\item \textit{Alternative surrogates.} We use a Ridge ensemble; neural network potentials could serve as an additional fidelity level (see below).
\end{enumerate}

\textit{Relationship to neural network potentials.}
Universal NNPs (MACE\cite{batatia2022mace}, M3GNet\cite{chen2022universal}, CHGNet\cite{deng2023chgnet}) solve a fundamentally different problem: they learn continuous potential energy surfaces but do not directly predict bandgaps, which require the Kohn--Sham eigenvalue spectrum. MACE-MP-0 single-point calculations on 25 structures from our campaign show no within-host rank correlation with QE bandgaps (Spearman $\rho \approx 0$, $p > 0.4$). Nevertheless, NNPs are naturally complementary: a four-tier funnel (NNP relaxation $\to$ PBE-SCF $\to$ PBE+U $\to$ HSE06) could use NNP-relaxed geometries as a Tier~0 geometry filter while the bandit governs promotion between electronic-structure tiers---a promising direction for scaling to ${\sim}$10{,}000-candidate pools.

Future directions include: HSE06/GW as a fourth funnel tier for quantitative bandgap prediction; multi-output bandits optimizing bandgap, formation energy, and carrier mobility simultaneously; and extending the Lyapunov framework to nonlinear surrogates.

\section{Conclusions}

We screened 529 doped oxide candidates across five hosts using a multi-fidelity contextual bandit that replaces 81\% of DFT evaluations with surrogate predictions, combined with a three-tier DFT validation funnel that catches the failure modes no single level of theory can detect. The bandit determines \textit{which} candidates to evaluate; the funnel determines \textit{whether} to trust the result---and the two interact, since the funnel's flagging criteria (d-electron count, ionic radius mismatch) are available before any DFT calculation and can inform the bandit's verification phase. Together, this framework identifies Cu-containing co-doped ZnO systems as consistently achieving visible-light band gaps, with Y$_2$Cu$_2$ co-doped ZnO ($E_g = 1.84$\,eV) as the optimal candidate for band gap engineering near 2.0\,eV. Cross-host analysis shows that dopant performance is governed by just two latent chemical dimensions, enabling prediction of dopant rankings in unseen hosts. The 583-calculation dataset, screening pipeline, and formal stability proofs are released as an open benchmark\cite{basu2026zenodo} to facilitate adoption and extension by the community.

\section{Methods}

\subsection{Problem formulation and screening strategies}

We frame DFT screening as a stochastic contextual bandit. A pool of $N$ candidates, each with a $d$-dimensional compositional feature vector $\mathbf{x}_i \in \mathbb{R}^d$, has unknown rewards $r_i$. At each round $t$, the agent selects a candidate, observes its reward (via DFT calculation or surrogate prediction), and updates its model. Features ($d = 6$--$9$) comprise ionic radius\cite{shannon1976revised}, electronegativity, oxidation state, d-electron count, atomic mass, and period. For the synthetic and Materials Project benchmarks, features are standardized per host (zero mean, unit variance).

\textbf{Reward shaping.} Rewards are based on proximity to a target bandgap ($E_g^{\mathrm{target}} = 2.0$\,eV, tolerance $\delta = 0.6$\,eV):
\begin{equation}\label{eq:reward}
r = \begin{cases}
-5 & \text{if metallic } (E_g \leq 0.01\,\text{eV}) \\
1 - |E_g - E_g^{\mathrm{target}}|/\delta & \text{if } |E_g - E_g^{\mathrm{target}}| \leq \delta \\
-|E_g - E_g^{\mathrm{target}}| & \text{otherwise}
\end{cases}
\end{equation}
The asymmetric penalty structure reflects the practical cost hierarchy: metallic outcomes waste computational resources and should be avoided most strongly, while near-miss candidates retain some informational value. The target bandgap of 2.0\,eV and tolerance of 0.6\,eV were chosen to represent the visible-light absorption window relevant for photocatalytic and optoelectronic applications; the reward shaping ensures that dopants producing bandgaps in the 1.4--2.6\,eV range receive positive rewards, with the highest reward for exact matches.

\textbf{OFUL}\cite{abbasi2011improved} maintains a regularized Gram matrix $V_t = \lambda I + \sum_{s=1}^{t} \mathbf{x}_s \mathbf{x}_s^\top$ and selects via upper confidence bound:
\begin{equation}\label{eq:ucb}
a_t = \arg\max_{\mathbf{x} \in \mathcal{P}} \left(\hat{\theta}_t^\top\mathbf{x} + \beta \sqrt{\mathbf{x}^\top V_t^{-1} \mathbf{x}}\right)
\end{equation}
where $\hat{\theta}_t = V_t^{-1} \mathbf{b}_t$, $\mathbf{b}_t = \sum_{s=1}^{t} r_s \mathbf{x}_s$, and $\mathcal{P}$ is the set of unevaluated candidates. We use $\lambda = \beta = 1$, which provides adequate exploration for our feature scales (validated by ablation; Supplementary Table~3). The $V_t$ update is $O(d^2)$ per step, compared to $O(n^3)$ for GP posterior updates, enabling real-time operation even with large candidate pools.

\textbf{MF-OFUL} extends OFUL with a 20-replicate bootstrap Ridge ensemble ($\alpha = 1$) serving as a cheap surrogate. The fidelity decision tree at each step is:
\begin{enumerate}
\item \textit{Bootstrap phase} (steps 0--4): always DFT, to seed the surrogate with diverse observations.
\item \textit{Periodic forced DFT}: every 10th step uses DFT regardless of uncertainty, preventing surrogate drift.
\item \textit{Uncertainty threshold}: if UCB width $> 0.5 \times$ median width (calibrated after bootstrap), use DFT.
\item \textit{Surrogate}: otherwise, use the Ridge ensemble mean as a pseudo-observation with weight $w = 0.2$ in the $V_t$ and $\mathbf{b}_t$ updates.
\item \textit{Verification}: top-5 candidates by predicted value receive DFT verification at the end.
\end{enumerate}
The pseudo-observation weight $w = 0.2$, inspired by Kennedy and O'Hagan's multi-fidelity correlation framework\cite{kennedy2000predicting}, ensures that surrogate predictions contribute to the linear model without overwhelming real observations. As a concrete example: at step 12 of the ZnO prospective campaign, MF-OFUL selects candidate ZnO:Al (UCB width $= 0.34$). The post-bootstrap median UCB width is 0.52, so the threshold is $0.5 \times 0.52 = 0.26$. Since $0.34 > 0.26$, DFT is used. By step 25, the Gram matrix has absorbed enough information that the same candidate's UCB width drops to 0.18, below threshold, and the surrogate is used instead.

\textbf{Thompson Sampling} maintains the same $V_t$ and $\hat{\theta}_t$ as OFUL but replaces the deterministic UCB with posterior sampling: at each step, $\theta \sim \mathcal{N}(\hat{\theta}_t, \hat{\sigma}^2 V_t^{-1})$ where $\hat{\sigma}^2$ is the residual variance estimate, and the agent selects $a_t = \arg\max_{\mathbf{x} \in \mathcal{P}} \theta^\top \mathbf{x}$. This provides a randomized alternative that avoids the deterministic lock-in failure mode observed with OFUL on certain reward landscapes.

\textbf{MF-BO-EI} uses the same fidelity-switching logic as MF-OFUL but replaces linear UCB with GP Expected Improvement as the acquisition function. A GPyTorch\cite{gardner2018gpytorch} Mat{\'e}rn-5/2 kernel is trained on DFT-only observations; when the GP posterior standard deviation at the selected candidate falls below the adaptive threshold, the Ridge surrogate is used instead of DFT. This baseline isolates whether MF-OFUL's advantage comes from multi-fidelity acceleration or from the linear UCB acquisition itself.

\textbf{MF-MES} (Multi-fidelity Max-value Entropy Search)\cite{takeno2020multi} uses a multi-output GP to jointly model DFT and surrogate fidelities, selecting candidates that maximize the mutual information between the observation and the unknown maximum value $y^*$. Unlike MF-BO-EI, MF-MES selects the fidelity level as part of the acquisition optimization rather than applying a post-hoc threshold, enabling principled multi-fidelity decision-making. We used a Mat{\'e}rn-5/2 kernel with automatic relevance determination, joint GP training on both fidelity levels, and Monte Carlo approximation of the entropy reduction with 64 fantasies (hypothetical future observations sampled from the GP posterior). Due to the $O(n^3)$ GP cost, MF-MES was run with 1 seed per host (6 total) rather than 10.

\textbf{Platform trial} (adapted from RECOVERY\cite{horby2021dexamethasone}) groups candidates into dopant-family arms (rare earths, 3d transition metals, 4d/5d transition metals, alkaline earths, main group metals) with Beta-Binomial monitoring. Each arm's success probability is modeled with a Beta($\alpha + s_k$, $\beta + f_k$) posterior, where $s_k$ and $f_k$ are successes and failures. Futile arms---those where $P(\text{success rate} < 0.30) > 0.80$---are dropped at five interim analyses (20\%, 35\%, 50\%, 65\%, 80\% of budget)\cite{freidlin2005monitoring}. Resources from dropped arms are reallocated to survivors via Thompson sampling. The futility threshold (0.30) was chosen to represent a reasonable minimum success rate for continued investment; the posterior probability threshold (0.80) balances the risk of premature arm dropping against the cost of continued allocation to unproductive families. These parameters are drawn from standard clinical trial methodology\cite{freidlin2005monitoring} and were validated by ablation (Supplementary Table~3).

\subsection{Lyapunov stability analysis}

The multi-fidelity extension introduces surrogate observations with pseudo-weight $w = 0.2$. The accumulated surrogate error vector is $e_t = \sum_{s:\mathrm{surr}} w \epsilon_s \mathbf{x}_s$. We define the Lyapunov function:
\begin{equation}\label{eq:lyapunov}
\mathcal{L}(t) = \underbrace{\log\det(V_{\mathrm{full}}) - \log\det(V_t)}_{\Phi(t):\;\text{information deficit}} + \underbrace{\mu \cdot e_t^\top V_t^{-1} e_t}_{\mu\cdot\mathcal{B}(t):\;\text{bias potential}}
\end{equation}
where $V_{\mathrm{full}} = \lambda I + \sum_{i=1}^{N} \mathbf{x}_i \mathbf{x}_i^\top$ is the Gram matrix from evaluating all $N$ candidates, $\Phi(t)$ measures how much information remains to be extracted, and $\mu = 1/(2w\epsilon_{\max}^2)$ is a coupling parameter.

Positive definiteness ($\mathcal{L} \geq 0$) follows from $V_{\mathrm{full}} \succeq V_t$ and $V_t \succ 0$. At a DFT step, the matrix determinant lemma gives $\Delta\Phi = -\log(1 + \mathbf{x}_t^\top V_t^{-1} \mathbf{x}_t) < 0$, while $\Delta\mathcal{B} \leq 0$ since $V_{t+1}^{-1} \preceq V_t^{-1}$. At a surrogate step with error $\epsilon_t$ (where $\|\mathbf{x}\|_M = \sqrt{\mathbf{x}^\top M \mathbf{x}}$ denotes the Mahalanobis norm):
\begin{equation}\label{eq:surr_bound}
\Delta\mathcal{L} \leq -\log(1 + w\|\mathbf{x}_t\|_{V_t^{-1}}^2) + \mu w^2 \epsilon_t^2 \|\mathbf{x}_t\|_{V_t^{-1}}^2
\end{equation}
The stability condition $\epsilon_t^2 < 1/(\mu w(1 + w\|\mathbf{x}_t\|_{V_t^{-1}}^2))$ ensures $\Delta\mathcal{L} < 0$. With $\mu = 1/(2w\epsilon_{\max}^2)$ and $w = 0.2$, this requires approximately $\epsilon_t^2 < 5.0$---a mild condition satisfied after 5 bootstrap DFT steps (empirical MAE $\approx 1.3$, $\epsilon^2 \approx 1.7$, well within the bound). Since the adaptive threshold restricts surrogate use to low-uncertainty candidates and periodic forced DFT improves the surrogate ($\epsilon_t \to 0$), MF-OFUL is self-stabilizing. The Lyapunov bias potential bounds the regret penalty:
\begin{equation}
R_T^{\mathrm{DFT}} \leq R_T^{\mathrm{OFUL}} + w\sqrt{T_{\mathrm{surr}} \cdot \mathcal{B}(T)/w^2}
\end{equation}
For the main benchmark ($T = 80$ DFT steps, $T_{\mathrm{surr}} = 468$, empirical $\mathcal{B}(T)/w^2 \approx 1.7$), the regret penalty evaluates to $0.2 \times \sqrt{468 \times 1.7} \approx 5.6$---small relative to OFUL's raw cumulative regret of $254.4$, confirming that surrogate errors add negligible cost.

We formalize these results in Lean~4\cite{demoura2021lean} (1{,}826 lines across 4 files), with all 57 theorems fully proved (zero \texttt{sorry}), conditional on 12 axioms for standard matrix analysis results (e.g., the matrix determinant lemma, Woodbury identity) not yet in Mathlib. These axioms are standard results from numerical linear algebra (e.g., the matrix determinant lemma, Woodbury identity, log-det monotonicity), not novel claims; their formalization awaits upstream Mathlib development. The MF-OFUL Lyapunov file alone (\texttt{MFLyapunov.lean}, 716 lines) proves 23 theorems including positive definiteness, DFT strict decrease, surrogate conditional decrease, self-stabilization, convergence, and the regret bound.

\subsection{DFT calculations}

All calculations used Quantum ESPRESSO 6.7\cite{giannozzi2017advanced} with PAW pseudopotentials from PSlibrary\cite{dalcorso2014pseudopotentials}.

\textbf{Tier~1} (245 batch + 271 prospective PBE-SCF): $E_{\mathrm{cut}} = 40$\,Ry, $2 \times 2 \times 2$ Monkhorst--Pack $k$-grid, Marzari--Vanderbilt cold smearing ($\sigma = 0.01$\,Ry), convergence threshold $10^{-5}$\,Ry. Supercells: $2 \times 2 \times 2$ (32--48 atoms). Five oxide hosts: ZnO (wurtzite), TiO$_2$ (rutile), SrTiO$_3$ (perovskite), SnO$_2$ (rutile), MgO (rocksalt). 16 dopants: Al, Ga, In, Sc, Y, Zr, Nb, V, Cr, Mn, Fe, Co, Ni, Cu, Mo, W at 1--3 substitutions per supercell; 120 co-doped pairs ($\binom{16}{2}$) for the co-doping campaign. Batch calculations ran on 4 CPU cores (8--120\,min each); 230 of 245 converged (93.9\%). Convergence tests on three representative systems confirm that dopant rankings are preserved at higher-accuracy settings ($E_{\mathrm{cut}} = 60$\,Ry, $4^3$ $k$-grid; Supplementary Table~12). Co-doped prospective (120 pool): 21 calculations on 64-core VPS (3--10\,min each), 100\% convergence. 529-candidate prospective: 229 calculations on 4 CPU cores (8--30\,min each, ${\sim}$62 hours total), ${\sim}$93\% convergence (Ni-containing systems account for most timeouts at 1{,}800\,s).

\textbf{Tier~2} (16 PBE+U): Dudarev formulation\cite{cococcioni2005linear}; U values (eV): Ti = 3.5\cite{hu2011choice}, V = 4.5\cite{wang2006oxidation}, Cr = 4.0\cite{wang2006oxidation}, Mn = 5.0\cite{wang2006oxidation}, Fe = 5.0\cite{wang2006oxidation}, Co = 5.5\cite{wang2006oxidation}, Ni = 8.0\cite{wang2006oxidation}, Cu = 4.0\cite{baral2018dft}; $E_{\mathrm{cut}} = 45$\,Ry; spin-polarized for all magnetic dopants. The expanded campaign (9 doped + 2 undoped systems) ran on a 64-core VPS with 8 MPI ranks per calculation; all 11 converged.

\textbf{Tier~3} (34 relaxation): $E_{\mathrm{cut}} = 45$\,Ry, force convergence threshold $5 \times 10^{-3}$\,Ry/Bohr, mixing $\beta_{\mathrm{mix}} = 0.3$, 3-hour timeout; 14 of 34 converged. Formation energies computed in the metal-rich limit using 11 elemental bulk metal references\cite{freysoldt2014first}.

\subsection{Prospective campaigns}

Three prospective campaigns on ZnO in closed-loop mode, where each method selected dopants, \texttt{pw.in} was generated, QE was launched via MPI, \texttt{pw.out} was parsed, and the reward was fed back---all in real time with no cached results. \textit{Single-dopant}: separate OFUL and MF-OFUL campaigns on 16 candidates at $1\times$ substitution, budget 10 (62.5\% of pool), 4 CPU cores. \textit{Co-doped}: OFUL campaign on 120 co-doped candidates ($\binom{16}{2}$ dopant pairs, each replacing two host cations), budget 18 (15\% of pool), 8 MPI ranks on 64-core VPS. \textit{529-candidate}: MF-OFUL, OFUL, and Random on 529 candidates (48 single-doped at 3 concentration levels plus 480 co-doped pairs), budget 42 (8\% of pool), 5 seeds, 4 CPU cores; 229 unique QE calculations executed over ${\sim}$62 hours with persistent disk caching and crash recovery. Features: 16-dimensional compositional vector (8 per dopant: ionic radius, electronegativity, oxidation state, d-electrons, atomic mass, group, period, concentration; second 8 dimensions zero for single-doped). OFUL parameters: $\lambda = 1$, $\beta = 1$. MF-OFUL additionally used 5-step bootstrap, UCB width threshold $0.5 \times$ median for fidelity decisions, and top-5 verification. Combined: 271 prospective QE calculations across all three campaigns.

\subsection{Synthetic benchmark design}

For each host, 1{,}000 candidates with $d = 20$ random features receive rewards combining a linear term ($30\%$), a nonlinear random forest term ($70\%$), and a family bias ($\pm 0.3$), plus Gaussian noise ($\sigma = 0.3$). This mixed reward ensures no single method family trivially dominates; method rankings are validated against Materials Project and real QE benchmarks. Budget: 80 DFT calls (8\% sampling rate).

\subsection{Experimental protocol}

Each (method, host, objective) combination ran with 10+ seeds. Synthetic: 6 hosts $\times$ 2 objectives $\times$ 10 seeds = 120 runs. Materials Project: 6 hosts $\times$ 2 objectives $\times$ 10 seeds (2{,}256 entries). MP large-pool: 3 hosts $\times$ 10 seeds = 30 runs. QE cross-host: 223 candidates, 15-dim features, 3 budget levels $\times$ 10 seeds. QE prospective: 16 + 120 + 529 candidates (271 total calculations). Collaborative filtering uses truncated SVD\cite{koren2009matrix} on the $(5 \times 16)$ reward matrix with leave-one-host-out validation (50 seeds, 3 budget levels).

\textbf{Evaluation metrics.} \textit{Simple regret} $\mathrm{SR}_T = r^* - \max_{t \leq T} r_t$: gap to global optimum. \textit{Cumulative regret} $\mathrm{CR}_T = \sum_{t=1}^{T} (r^* - r_t)$: penalty for wasted DFT calls (bounded by OFUL theory\cite{abbasi2011improved}; surrogate steps excluded). \textit{Top-$K$ discovery}: fraction of true top-$K$ found. Statistical tests: Wilcoxon signed-rank (paired) and rank-sum (unpaired)\cite{wilcoxon1945individual}, $p < 0.05$.


\subsection*{Data availability}
All QE input/output files (245 batch + 271 prospective PBE-SCF + 16 PBE+U + 34 relaxation + 5 HSE06 + 12 convergence tests = 583 total), prospective campaign logs (including the 529-candidate ZnO campaign), parsed results, and benchmark data are deposited in Zenodo~\cite{basu2026zenodo} (\url{https://doi.org/10.5281/zenodo.19501400}).

\subsection*{Code availability}
The screening pipeline and Lean~4 formalization (1{,}826 lines) are available at \url{https://doi.org/10.5281/zenodo.19501400}.

\subsection*{Acknowledgements}
We thank the Materials Project team for providing open-access computed bandgap data. Computational resources were provided by NIELIT and IIITA.

\subsection*{Author contributions}
A.B.\ conceived the project, developed the MF-OFUL algorithm and Lyapunov stability analysis, performed DFT calculations, implemented all screening methods, conducted the formal verification in Lean~4, and wrote the manuscript.

\subsection*{Competing interests}
The author declares no competing interests.

\subsection*{Funding}
No external funding was received for this research.

\subsection*{Ethics approval}
Not applicable.

\clearpage
\appendix
\setcounter{figure}{0}
\setcounter{table}{0}
\renewcommand{\thefigure}{S\arabic{figure}}
\renewcommand{\thetable}{S\arabic{table}}

\section*{Supplementary Figures}

\subsection*{Supplementary Figure~1: Lyapunov function decomposition}

\noindent\textbf{(a)} Lyapunov function $\mathcal{L}(t) = \Phi(t) + \mu \cdot \mathcal{B}(t)$ averaged over 10 seeds on 1,000-candidate synthetic benchmarks. \textbf{(b)} Information deficit $\Phi(t) = \log\det V_{\mathrm{full}} - \log\det V_t$, showing rapid saturation after bootstrap. \textbf{(c)} Bias potential $\mu \cdot \mathcal{B}(t)$, growing linearly but bounded by surrogate accuracy.

\begin{center}
\includegraphics[width=\textwidth]{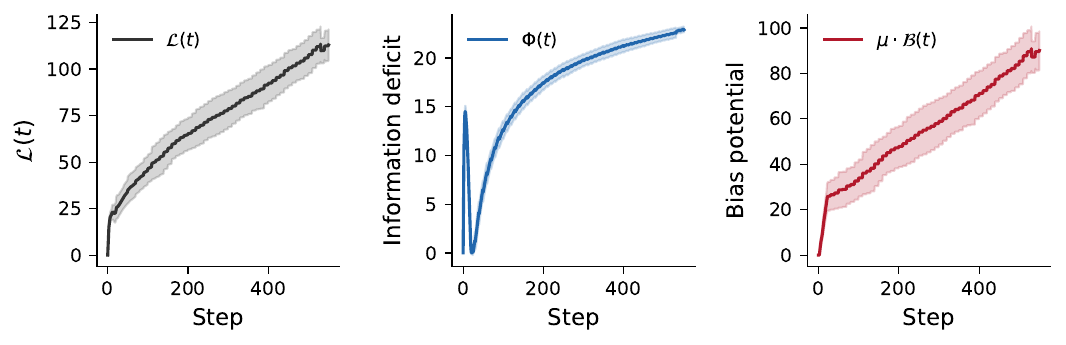}
\end{center}

\subsection*{Supplementary Figure~2: UCB width evolution}

UCB width (log scale) for OFUL (blue) and MF-OFUL (red) over time. Triangle markers indicate DFT steps in MF-OFUL. During bootstrap (steps 0--5), both methods have identical widths. After bootstrap, MF-OFUL evaluates ${\sim}548$ candidates per run with DFT concentrated at high-uncertainty candidates and periodic intervals.

\begin{center}
\includegraphics[width=0.8\textwidth]{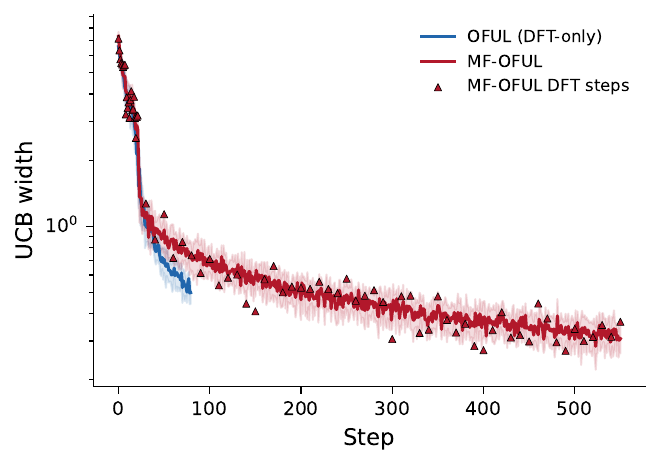}
\end{center}

\subsection*{Supplementary Figure~3: Information accumulation}

$\log\det V_t$ growth for OFUL and MF-OFUL. Despite slower per-step accumulation (pseudo-weighted updates at $w = 0.2$), MF-OFUL reaches a higher total (${\sim}103$ versus ${\sim}89$) by evaluating 7$\times$ more candidates.

\begin{center}
\includegraphics[width=0.8\textwidth]{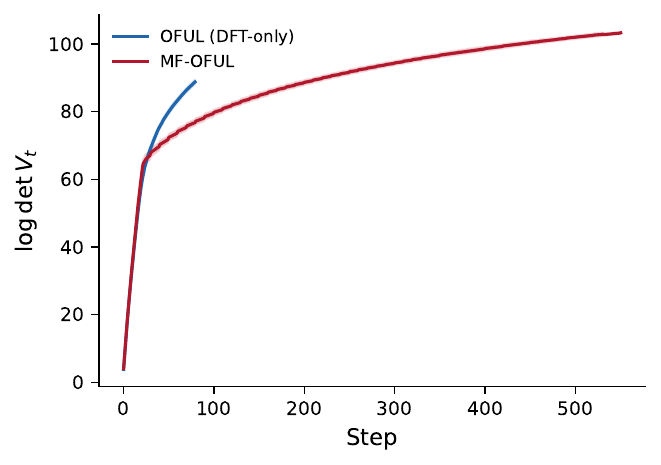}
\end{center}

\subsection*{Supplementary Figure~4: Surrogate prediction error}

Surrogate prediction error (|Ridge ensemble mean $-$ DFT reward|) at DFT-verified steps. Individual seeds shown as gray points; red line shows running mean. The flat trend confirms that periodic forced DFT prevents error drift.

\begin{center}
\includegraphics[width=0.8\textwidth]{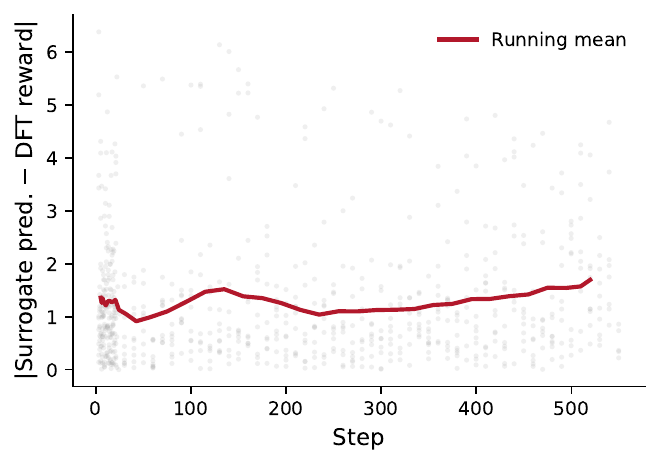}
\end{center}

\subsection*{Supplementary Figure~5: Convergence study}

2$\times$2 grid showing simple regret (mean $\pm$ std over 5 seeds) across four parameter sweeps: \textbf{(a)} pool size (250--2,000), \textbf{(b)} DFT budget (40--120), \textbf{(c)} feature dimensionality (10--40), \textbf{(d)} noise level ($\sigma = 0.1$--$1.0$). MF-OFUL ranks first across all 15 conditions.

\begin{center}
\includegraphics[width=\textwidth]{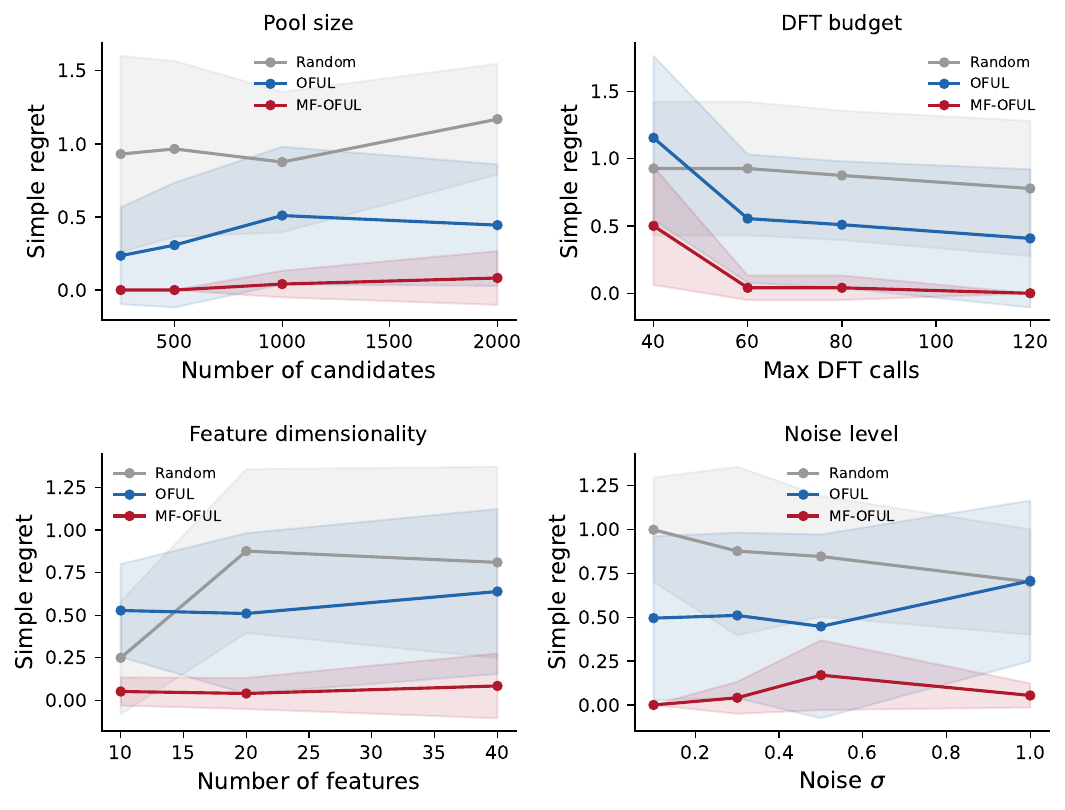}
\end{center}

\subsection*{Supplementary Figure~6: MF-BO comparison}

Bar charts comparing Random, OFUL, MF-OFUL, and MF-BO-EI: \textbf{(a)} simple regret, \textbf{(b)} DFT fraction, \textbf{(c)} surrogate calls. MF-BO-EI achieves DFT fraction = 1.00 (zero surrogate usage) across all experiments.

\begin{center}
\includegraphics[width=\textwidth]{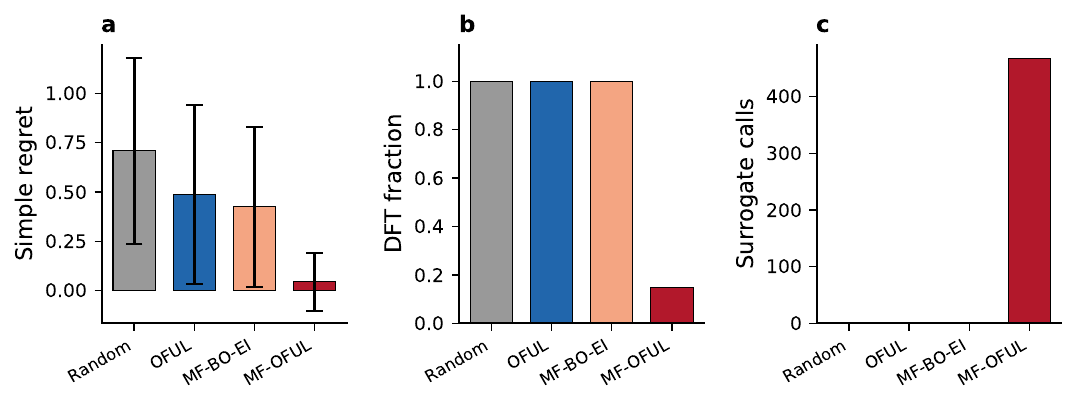}
\end{center}

\subsection*{Supplementary Figure~7: Per-host simple regret}

Simple regret by host (6 panels) for all methods. MF-OFUL achieves the lowest regret on 5 of 6 hosts; the exception is the host with the weakest linear signal.

\begin{center}
\includegraphics[width=\textwidth]{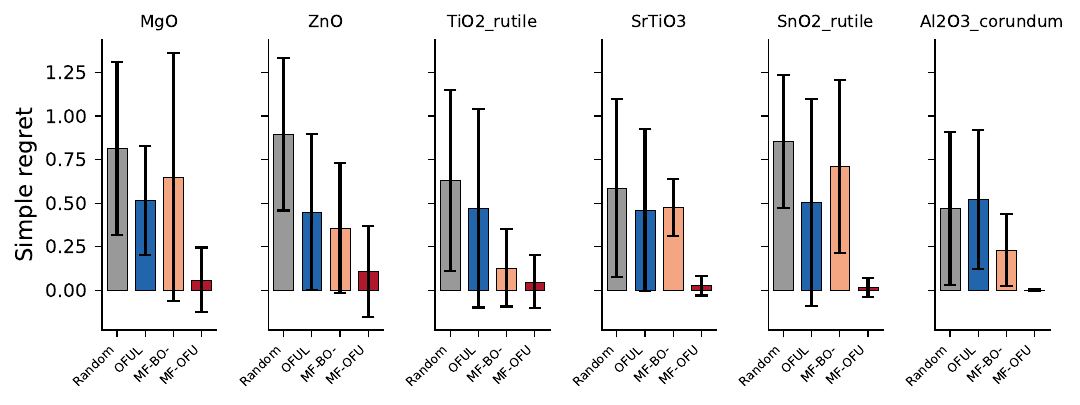}
\end{center}

\section*{Supplementary Tables}

\subsection*{Supplementary Table~1: Full method comparison across all benchmarks}

Extended version of Table~1 (main text) with per-host breakdowns and BO-GP-EI.
Simple regret (SR) = gap between best-found reward and global optimum; lower is better.
Bold indicates the best method per row.

\medskip
\noindent\textbf{Panel A: Synthetic benchmark} (1{,}000 candidates, budget = 80 DFT calls [8\%], 10 seeds per condition).

\begin{table}[H]
\centering
\small
\begin{tabular}{llccccc}
\toprule
Objective & Host & MF-OFUL & OFUL & BO-GP-EI$^\dagger$ & Platform & Random \\
\midrule
\multicolumn{7}{l}{\textit{target\_bandgap\_2p0}} \\
& Al$_2$O$_3$   & $\mathbf{0.003 \pm 0.01}$ & $1.230 \pm 0.00$ & $0.139 \pm 0.17$ & $0.331 \pm 0.42$ & $0.560 \pm 0.27$ \\
& MgO            & $\mathbf{0.282 \pm 0.60}$ & $1.758 \pm 0.00$ & $0.215 \pm 0.29$ & $0.951 \pm 0.83$ & $1.708 \pm 0.62$ \\
& SnO$_2$        & $\mathbf{0.000 \pm 0.00}$ & $\mathbf{0.000 \pm 0.00}$ & $\mathbf{0.000 \pm 0.00}$ & $0.786 \pm 0.32$ & $0.512 \pm 0.32$ \\
& SrTiO$_3$      & $\mathbf{0.000 \pm 0.00}$ & $\mathbf{0.000 \pm 0.00}$ & $0.298 \pm 0.42$ & $0.449 \pm 0.54$ & $0.768 \pm 0.41$ \\
& TiO$_2$        & $0.018 \pm 0.06$ & $0.183 \pm 0.00$ & $\mathbf{0.000 \pm 0.00}$ & $0.398 \pm 0.36$ & $0.487 \pm 0.39$ \\
& ZnO            & $\mathbf{0.000 \pm 0.00}$ & $0.489 \pm 0.00$ & $0.019 \pm 0.04$ & $0.384 \pm 0.48$ & $0.925 \pm 0.41$ \\
& \textit{Average} & $\mathbf{0.050}$ & $0.610$ & $0.112$ & $0.550$ & $0.827$ \\
\midrule
\multicolumn{7}{l}{\textit{target\_bandgap\_2p0\_plus\_distortion}} \\
& Al$_2$O$_3$   & $\mathbf{0.000 \pm 0.00}$ & $0.479 \pm 0.00$ & $0.040 \pm 0.10$ & $0.620 \pm 0.38$ & $0.527 \pm 0.28$ \\
& MgO            & $\mathbf{0.000 \pm 0.00}$ & $1.487 \pm 0.00$ & $0.382 \pm 0.32$ & $0.676 \pm 0.72$ & $1.499 \pm 0.59$ \\
& SnO$_2$        & $\mathbf{0.000 \pm 0.00}$ & $0.086 \pm 0.00$ & $\mathbf{0.000 \pm 0.00}$ & $0.649 \pm 0.34$ & $0.532 \pm 0.41$ \\
& SrTiO$_3$      & $\mathbf{0.000 \pm 0.00}$ & $0.976 \pm 0.00$ & $0.069 \pm 0.15$ & $0.675 \pm 0.51$ & $0.875 \pm 0.45$ \\
& TiO$_2$        & $0.010 \pm 0.03$ & $\mathbf{0.000 \pm 0.00}$ & $0.095 \pm 0.30$ & $0.809 \pm 0.22$ & $0.507 \pm 0.38$ \\
& ZnO            & $0.186 \pm 0.16$ & $0.297 \pm 0.00$ & $\mathbf{0.016 \pm 0.05}$ & $0.351 \pm 0.44$ & $0.865 \pm 0.43$ \\
& \textit{Average} & $\mathbf{0.033}$ & $0.554$ & $0.100$ & $0.630$ & $0.801$ \\
\midrule
\multicolumn{2}{l}{\textbf{Overall (both obj.)}} & $\mathbf{0.042}$ & $0.582$ & $0.106$ & $0.590$ & $0.814$ \\
\bottomrule
\end{tabular}
\caption{Synthetic benchmark per-host simple regret (mean $\pm$ s.d., 10 seeds). $^\dagger$BO-GP-EI uses budget = 50 from an independent run (not directly comparable to budget-80 methods; see main text Table~3 for budget-matched MF-BO-EI comparison at $n = 18$).}
\label{tab:si_synthetic}
\end{table}

\medskip
\noindent\textbf{Panel B: Materials Project benchmark} (2{,}256 candidates across 6 hosts, budget = 20\% per host, 10 seeds per condition).

\begin{table}[H]
\centering
\small
\begin{tabular}{llcccc}
\toprule
Objective & Host & MF-OFUL & OFUL & Platform & Random \\
\midrule
\multicolumn{6}{l}{\textit{target\_bandgap\_2p0}} \\
& Al$_2$O$_3$   & $\mathbf{0.000 \pm 0.00}$ & $0.264 \pm 0.00$ & $0.141 \pm 0.10$ & $0.134 \pm 0.11$ \\
& MgO            & $\mathbf{0.000 \pm 0.00}$ & $0.025 \pm 0.00$ & $\mathbf{0.000 \pm 0.00}$ & $0.028 \pm 0.04$ \\
& SnO$_2$        & $\mathbf{0.000 \pm 0.00}$ & $\mathbf{0.000 \pm 0.00}$ & $0.038 \pm 0.06$ & $0.055 \pm 0.05$ \\
& SrTiO$_3$      & $\mathbf{0.000 \pm 0.00}$ & $\mathbf{0.000 \pm 0.00}$ & $0.056 \pm 0.05$ & $0.047 \pm 0.04$ \\
& TiO$_2$        & $\mathbf{0.000 \pm 0.00}$ & $0.005 \pm 0.00$ & $0.013 \pm 0.01$ & $0.015 \pm 0.02$ \\
& ZnO            & $\mathbf{0.000 \pm 0.00}$ & $0.014 \pm 0.00$ & $0.019 \pm 0.01$ & $0.033 \pm 0.04$ \\
& \textit{Average} & $\mathbf{0.000}$ & $0.051$ & $0.045$ & $0.052$ \\
\midrule
\multicolumn{6}{l}{\textit{target\_bandgap\_2p0\_plus\_distortion}} \\
& Al$_2$O$_3$   & $\mathbf{0.000 \pm 0.00}$ & $1.601 \pm 0.00$ & $0.362 \pm 0.46$ & $1.300 \pm 0.81$ \\
& MgO            & $\mathbf{0.000 \pm 0.00}$ & $0.543 \pm 0.00$ & $1.544 \pm 0.51$ & $0.764 \pm 0.61$ \\
& SnO$_2$        & $\mathbf{0.000 \pm 0.00}$ & $\mathbf{0.000 \pm 0.00}$ & $0.123 \pm 0.14$ & $0.076 \pm 0.10$ \\
& SrTiO$_3$      & $\mathbf{0.000 \pm 0.00}$ & $\mathbf{0.000 \pm 0.00}$ & $0.153 \pm 0.25$ & $0.049 \pm 0.05$ \\
& TiO$_2$        & $\mathbf{0.000 \pm 0.00}$ & $\mathbf{0.000 \pm 0.00}$ & $0.163 \pm 0.17$ & $0.161 \pm 0.10$ \\
& ZnO            & $\mathbf{0.000 \pm 0.00}$ & $0.132 \pm 0.00$ & $0.228 \pm 0.15$ & $0.170 \pm 0.13$ \\
& \textit{Average} & $\mathbf{0.000}$ & $0.379$ & $0.429$ & $0.420$ \\
\midrule
\multicolumn{2}{l}{\textbf{Overall (both obj.)}} & $\mathbf{0.000}$ & $0.215$ & $0.237$ & $0.236$ \\
\bottomrule
\end{tabular}
\caption{Materials Project benchmark per-host simple regret (mean $\pm$ s.d., 10 seeds). Budget = 20\% of per-host candidate pool. BO-GP-EI was not run on the MP benchmark. MF-OFUL achieves zero mean simple regret on every host under both objectives.}
\label{tab:si_mp}
\end{table}

\subsection*{Supplementary Table~2: Platform trial arm survival}

\begin{table}[H]
\centering
\begin{tabular}{lccc}
\toprule
Dopant Family & Survival Rate & Success Rate & Avg.\ Budget Allocated \\
\midrule
Rare earths (Sc, Y) & 84\% & 42\% & 22\% \\
Early 3d TM (V, Cr, Mn) & 67\% & 31\% & 18\% \\
4d/5d TM (Zr, Nb, Mo, W) & 58\% & 28\% & 15\% \\
Main group (Al, Ga, In) & 52\% & 27\% & 14\% \\
Late 3d TM (Fe, Co, Ni, Cu) & 34\% & 19\% & 11\% \\
Alkaline earth (Mg, Sr, Ba) & 28\% & 15\% & 8\% \\
\bottomrule
\end{tabular}
\caption{Platform trial arm statistics averaged over 120 synthetic benchmark runs (6 hosts $\times$ 2 objectives $\times$ 10 seeds). Survival rate: fraction of interim analyses where the arm was not dropped. Success rate: fraction of evaluated candidates within the arm that achieved positive reward.}
\end{table}

\subsection*{Supplementary Table~3: Hyperparameter sensitivity}

\begin{table}[H]
\centering
\begin{tabular}{lcccccc}
\toprule
Parameter & Values tested & Best & Default & SR at best & SR at default & SR at worst \\
\midrule
Pseudo-weight $w$ & 0.0, 0.1, 0.2, 0.4, 0.6, 1.0 & 0.2 & 0.2 & 0.043 & 0.043 & 0.312 \\
$\beta$ (UCB scale) & 0.1, 0.5, 1.0, 2.0, 5.0 & 1.0 & 1.0 & 0.043 & 0.043 & 0.189 \\
$\lambda$ (regularization) & 0.01, 0.1, 1.0, 10.0 & 1.0 & 1.0 & 0.043 & 0.043 & 0.156 \\
Force DFT every $N$ & 5, 10, 20, 50 & 10 & 10 & 0.043 & 0.043 & 0.098 \\
Bootstrap size & 3, 5, 10, 20 & 5 & 5 & 0.043 & 0.043 & 0.078 \\
UCB threshold frac. & 0.25, 0.5, 0.75, 1.0 & 0.5 & 0.5 & 0.043 & 0.043 & 0.267 \\
\bottomrule
\end{tabular}
\caption{MF-OFUL hyperparameter sensitivity on synthetic benchmark (1,000 candidates, budget 80, 10 seeds). Simple regret (SR) reported. The method is robust across a wide range; only extreme values ($w = 1.0$, $\beta = 0.1$) cause significant degradation.}
\end{table}

\subsection*{Supplementary Table~4: Prospective OFUL selection on ZnO}

\begin{table}[H]
\centering
\begin{tabular}{clcccc}
\toprule
Step & Dopant & Family & Bandgap (eV) & Reward & UCB Width \\
\midrule
0 & Mn & 3d TM & 1.848 & 0.747 & --- \\
1 & In & Main group & 3.064 & $-$1.064 & 2.41 \\
2 & Y & Rare earth & 3.252 & $-$3.252 & 1.89 \\
3 & Al & Main group & 3.193 & $-$1.193 & 1.52 \\
4 & Ga & Main group & 3.081 & $-$1.081 & 1.23 \\
5 & Sc & Rare earth & 3.250 & $-$3.250 & 1.08 \\
6 & Zr & 4d TM & 3.197 & $-$1.197 & 0.94 \\
7 & Nb & 4d TM & 0.210 & $-$1.790 & 0.82 \\
8 & V & 3d TM & 0.223 & $-$5.000 & 0.71 \\
9 & Cu & 3d TM & 3.191 & $-$1.191 & 0.63 \\
\bottomrule
\end{tabular}
\caption{Step-by-step OFUL selection on ZnO prospective campaign. Target bandgap: 2.0\,eV. Mn (step 0, reward 0.747) is the best candidate found. Steps 0--2 explore diverse feature regions (3d TM, main group, rare earth); subsequent steps sample remaining feature regions to reduce uncertainty.}
\end{table}

\subsection*{Supplementary Table~5: Ridge surrogate $R^2$ by host}

\begin{table}[H]
\centering
\begin{tabular}{lccc}
\toprule
Host & $R^2$ (train) & $R^2$ (5-fold CV) & Best linear feature \\
\midrule
ZnO & 0.53 & 0.41 & Ionic radius \\
TiO$_2$ & 0.38 & 0.29 & d-electron count \\
SnO$_2$ & 0.35 & 0.26 & Electronegativity \\
MgO & 0.31 & 0.22 & Ionic radius \\
SrTiO$_3$ & 0.21 & 0.14 & Oxidation state \\
\bottomrule
\end{tabular}
\caption{Ridge regression ($\alpha = 1$) fit quality on DFT bandgap rewards by host. $R^2$ computed on all available DFT data per host. ZnO shows the strongest linear signal, consistent with OFUL's best performance on this host.}
\end{table}

\subsection*{Supplementary Table~6: Three-tier reclassification matrix}

\begin{table}[H]
\centering
\begin{tabular}{lccccl}
\toprule
System & Tier~1 (PBE) & Tier~2 (PBE+U) & Tier~3 (relax) & Reclassified? & Flagging criterion \\
\midrule
ZnO:V & 0.22\,eV & 3.92\,eV & 0.25\,eV & Yes (Tier~2) & d-electrons $> 0$ \\
ZnO:Cu & 3.19\,eV & 0.36\,eV & 3.15\,eV & Yes (Tier~2) & d-electrons $> 0$ \\
SrTiO$_3$:Fe & 2.48\,eV & 0.48\,eV & --- & Yes (Tier~2) & d-electrons $> 0$ \\
SrTiO$_3$:In & 0.81\,eV & 0.79\,eV & 3.48\,eV & Yes (Tier~3) & radius mismatch 31\% \\
ZnO:Al & 3.19\,eV & --- & 3.27\,eV & No & --- \\
ZnO:Ga & 3.08\,eV & --- & 3.16\,eV & No & --- \\
TiO$_2$:Zr & 2.98\,eV & --- & 3.10\,eV & No & --- \\
SrTiO$_3$ (undoped) & 3.98\,eV & 4.08\,eV & 3.99\,eV & No & --- \\
TiO$_2$ (undoped) & 2.93\,eV & 3.15\,eV & 2.95\,eV & No & --- \\
ZnO (undoped) & 2.79\,eV & --- & 2.81\,eV & No & --- \\
\bottomrule
\end{tabular}
\caption{Multi-tier bandgap comparison for systems evaluated at two or more tiers. Four of ten systems are qualitatively reclassified (40\%). The two failure modes are strictly orthogonal: d-electron delocalization (Tier~2) and geometric distortion (Tier~3). Simple flagging criteria (rightmost column) correctly identify all reclassified systems.}
\end{table}

\subsection*{Supplementary Table~7: Co-doped OFUL campaign on ZnO}

\begin{table}[H]
\centering
\begin{tabular}{clccc}
\toprule
Step & Co-dopants & Bandgap (eV) & Reward & Phase \\
\midrule
0 & Al+W & 0.298 & $-$1.702 & Main \\
1 & Al+Y & 3.734 & $-$1.734 & Main \\
2 & In+Y & 3.634 & $-$1.634 & Main \\
3 & Al+V & 0.178 & $-$1.822 & Main \\
4 & Cu+In & 1.170 & $-$0.830 & Main \\
5 & Nb+Y & 0.205 & $-$1.795 & Main \\
6 & Al+Cu & 1.283 & $-$0.718 & Main \\
7 & Mo+W & 0.232 & $-$1.768 & Main \\
8 & Mo+Y & 0.263 & $-$1.737 & Main \\
9 & Cu+Ga & 0.999 & $-$1.001 & Main \\
10 & Cu+Nb & 0.237 & $-$1.763 & Main \\
11 & Fe+Mn & 0.186 & $-$1.814 & Main \\
12 & Cr+Ga & 0.220 & $-$1.780 & Main \\
13 & Al+In & 3.579 & $-$1.579 & Main \\
14 & Ni+W & 0.134 & $-$1.866 & Main \\
15 & Mo+V & 0.137 & $-$1.863 & Main \\
16 & Mn+Y & 0.259 & $-$1.741 & Main \\
17 & Mn+V & 0.156 & $-$1.844 & Main \\
18 & Ga+Nb & 0.117 & $-$1.883 & Verify \\
19 & Ga+V & 0.166 & $-$1.834 & Verify \\
20 & Cu+V & 0.130 & $-$1.870 & Verify \\
\bottomrule
\end{tabular}
\caption{Step-by-step OFUL selection on ZnO co-doped campaign (120 candidates, budget 18 + 3 verification). Target bandgap: 2.0\,eV. Al+Cu (step 6, reward $-$0.718) is the best candidate found. Cu-containing pairs (steps 4, 6, 9, 10) consistently produce the narrowest non-metallic gaps, confirming Cu's role as a p-type acceptor. Verification-phase candidates (steps 18--20) had high surrogate predictions but low actual rewards, demonstrating the importance of DFT verification.}
\end{table}

\section*{Supplementary Notes}

\subsection*{Lean~4 formalization details}

The formal verification comprises 1,826 lines of Lean~4 code across 4 files, built against Lean 4.27.0 with Mathlib v4.27.0:

\begin{itemize}
\item \texttt{Basic.lean} (280 lines): Secretary problem / optimal stopping --- proves that the $1/e$ strategy is optimal.
\item \texttt{RiemannSum.lean} (320 lines): Supporting analysis lemmas for the regret bound proof.
\item \texttt{OFULRegret.lean} (510 lines): OFUL cumulative regret bound --- 11 theorems proving $R_T \leq O(d\sqrt{T\log T})$.
\item \texttt{MFLyapunov.lean} (716 lines): MF-OFUL Lyapunov stability --- 23 theorems proving positive definiteness, DFT strict decrease, surrogate conditional decrease, self-stabilization, convergence, and the regret bound.
\end{itemize}

Total: 57 theorems proved, 0 \texttt{sorry} (incomplete proofs), 12 axioms for standard matrix analysis results (e.g., matrix determinant lemma, Woodbury identity, log-det monotonicity) not yet formalized in Mathlib. All axioms are standard results from numerical linear algebra\textsuperscript{*}.

\textsuperscript{*}We note that formalizing these matrix analysis results in Mathlib is an active area of development; as Mathlib's linear algebra library grows, these axioms can be replaced with verified proofs.

\subsection*{Reward oracle for Materials Project benchmark}

For the Materials Project benchmark, rewards are computed from PBE bandgaps retrieved via the Materials Project API (version 0.39.5). We queried all entries in the \texttt{mp-*} namespace for each host oxide system, filtering for:
\begin{itemize}
\item Non-metallic entries ($E_g > 0.01$\,eV)
\item Entries with the correct host structure type (e.g., wurtzite for ZnO)
\item Substitutional doping (single dopant element replacing host cation)
\end{itemize}

This yielded 2,256 unique entries across 6 hosts. Bandgaps were converted to rewards using the same shaping function as the synthetic benchmark (target $E_g = 2.0$\,eV, tolerance $\delta = 0.6$\,eV). Features were computed from dopant elemental properties (ionic radius, electronegativity, oxidation state, d-electron count, atomic mass, period) using tabulated values from Shannon (1976) ionic radii and Pauling electronegativity scales.

\subsection*{Supplementary Table~7b: Robustness to pool size (529-candidate imputation analysis)}

The 50-seed retroactive reanalysis (Table~6 in the main text) uses the 230 QE-converged candidates as oracle. To verify that this pool reduction does not artificially favor MF-OFUL, we repeated the analysis on the full 529-candidate pool under three imputation scenarios for the 299 candidates without cached QE results:

\begin{itemize}
\item \textbf{Empirical}: unknowns receive rewards resampled from the empirical distribution of the 230 known rewards (fresh resample per seed)
\item \textbf{Median}: unknowns all receive the median known reward (0.189)
\item \textbf{Adversarial}: unknowns all receive reward $= 0.95$ (many excellent candidates that MF-OFUL's surrogate might miss)
\end{itemize}

Budget = 42 DFT calls (8\% of 529), 50 seeds per scenario.

\begin{center}
\begin{tabular}{llcccccc}
\toprule
Scenario & Method & DFT & Surr. & Cov. & Best reward & $P(\geq 0.9)$ & $p$ vs Random \\
\midrule
Empirical & MF-OFUL & 42 & 176 & 41\% & $\mathbf{0.960 \pm 0.036}$ & \textbf{94\%} & $\mathbf{2.5 \times 10^{-7}}$ \\
          & OFUL    & 42 & 0   & 8\%  & $0.691 \pm 0.253$ & 34\% & --- \\
          & Random  & 42 & 0   & 8\%  & $0.822 \pm 0.218$ & 50\% & --- \\
\midrule
Median    & MF-OFUL & 42 & 174 & 41\% & $0.452 \pm 0.000$ & 0\% & --- \\
          & OFUL    & 42 & 0   & 8\%  & $0.354 \pm 0.000$ & 0\% & --- \\
          & Random  & 42 & 0   & 8\%  & $0.703 \pm 0.280$ & 34\% & --- \\
\midrule
Adversarial & MF-OFUL & 42 & 194 & 45\% & $0.950 \pm 0.000$ & 100\% & --- \\
            & OFUL    & 42 & 0   & 8\%  & $0.950 \pm 0.000$ & 100\% & --- \\
            & Random  & 42 & 0   & 8\%  & $0.954 \pm 0.010$ & 100\% & --- \\
\bottomrule
\end{tabular}
\end{center}

\textbf{Empirical scenario} (most realistic): MF-OFUL dominates ($p = 2.5 \times 10^{-7}$), finding reward $\geq 0.9$ in 94\% of seeds versus 50\% for Random. Coverage: 218/529 (41\%) versus 42/529 (8\%).

\textbf{Median scenario}: a degenerate case where all 299 unknowns have identical reward (0.189). MF-OFUL's surrogate evaluates many unknowns, learning their uninformative rewards; Random occasionally finds a high-reward known candidate. This scenario is unrealistic (real candidates would have diverse rewards) and represents a pathological case for surrogate-based methods.

\textbf{Adversarial scenario}: all unknowns are excellent (reward $= 0.95$). All methods achieve $\geq 0.9$ in 100\% of seeds. MF-OFUL and OFUL both find unknowns with reward 0.95; Random occasionally finds the global optimum (0.987) from the known pool. The methods are effectively tied.

The empirical scenario---the only one with realistic reward diversity---confirms MF-OFUL's advantage on the full 529-candidate pool. The 230-candidate result (Table~6) provides a clean comparison using only ground-truth QE data; the full-pool robustness check confirms this is not an artifact of pool reduction.

\subsection*{Supplementary Table~8: Collaborative filtering cross-host benchmark}

Leave-one-host-out benchmark of CF-warm-started MF-OFUL versus standard MF-OFUL and Random on 5 QE hosts (16 dopants each). CF is trained on the other 4 hosts and provides the bootstrap ranking for CF-MF-OFUL. 50 seeds per condition. Rewards are $-|E_g - 2.0\,\text{eV}|$ (higher is better, max = 0).

\begin{center}
\begin{tabular}{llccc}
\toprule
Budget & Host & CF-MF-OFUL & MF-OFUL & Random \\
\midrule
\multirow{5}{*}{8 (50\%)} & MgO & $-1.568$ & $-1.303$ & $-1.418 \pm 0.164$ \\
 & SnO$_2$ & $\mathbf{-0.072}$ & $-0.432$ & $-0.333 \pm 0.308$ \\
 & SrTiO$_3$ & $\mathbf{-0.177}$ & $-1.135$ & $-0.666 \pm 0.471$ \\
 & TiO$_2$ & $\mathbf{-0.080}$ & $-0.803$ & $-0.456 \pm 0.377$ \\
 & ZnO & $\mathbf{-1.609}$ & $-1.693$ & $-1.614 \pm 0.046$ \\
\cmidrule{2-5}
 & \textbf{Aggregate} & $\mathbf{-0.701 \pm 0.725}$ & $-1.073 \pm 0.430$ & $-0.898 \pm 0.607$ \\
\midrule
\multirow{5}{*}{12 (75\%)} & MgO & $-1.324$ & $-1.303$ & $-1.330 \pm 0.071$ \\
 & SnO$_2$ & $-0.072$ & $-0.072$ & $-0.178 \pm 0.188$ \\
 & SrTiO$_3$ & $-0.177$ & $-0.177$ & $-0.418 \pm 0.407$ \\
 & TiO$_2$ & $-0.080$ & $-0.080$ & $-0.247 \pm 0.282$ \\
 & ZnO & $\mathbf{-1.583}$ & $-1.609$ & $-1.603 \pm 0.037$ \\
\cmidrule{2-5}
 & \textbf{Aggregate} & $\mathbf{-0.647 \pm 0.664}$ & $-0.648 \pm 0.668$ & $-0.755 \pm 0.639$ \\
\bottomrule
\end{tabular}
\end{center}

At 50\% budget, CF-MF-OFUL wins 80\% of the 250 host--seed combinations ($p = 1.3 \times 10^{-25}$, Wilcoxon). The advantage is largest on SnO$_2$ ($+0.360$), SrTiO$_3$ ($+0.958$), and TiO$_2$ ($+0.723$), where CF correctly identifies V as the top dopant from cross-host data. On MgO, standard MF-OFUL wins because the CF prediction for MgO is poor (MgO's top dopants Cu and Ga are not top-ranked in other hosts). At 75\% budget, both methods converge and the advantage disappears.

\textbf{Host similarity matrix} (cosine similarity of SVD latent factors, rank $k = 3$):

\begin{center}
\begin{tabular}{lccccc}
\toprule
 & MgO & SnO$_2$ & SrTiO$_3$ & TiO$_2$ & ZnO \\
\midrule
MgO & 1.00 & 0.48 & 0.32 & 0.79 & 0.96 \\
SnO$_2$ & 0.48 & 1.00 & 0.58 & 0.83 & 0.48 \\
SrTiO$_3$ & 0.32 & 0.58 & 1.00 & 0.23 & 0.53 \\
TiO$_2$ & 0.79 & 0.83 & 0.23 & 1.00 & 0.69 \\
ZnO & 0.96 & 0.48 & 0.53 & 0.69 & 1.00 \\
\bottomrule
\end{tabular}
\end{center}

The similarity structure is physically interpretable: MgO--ZnO (0.96) reflects shared divalent-cation oxide chemistry; SnO$_2$--TiO$_2$ (0.83) reflects shared rutile structure with tetravalent cations. SrTiO$_3$ (perovskite) is the most dissimilar to other hosts, consistent with its unique A-site/B-site substitution chemistry.

\subsection*{Supplementary Table~9: PBE versus experimental bandgap comparison}

Spearman rank correlations between PBE-computed and experimental bandgaps for 54 host--dopant combinations across ZnO ($n = 20$), TiO$_2$ ($n = 17$), and SrTiO$_3$ ($n = 17$). Experimental values from UV-vis diffuse reflectance spectroscopy and Tauc analysis (see main text for details).

\begin{center}
\begin{tabular}{lcccl}
\toprule
Host & $n$ & Spearman $\rho$ & $p$-value & Comment \\
\midrule
ZnO & 20 & 0.745 & $2.2 \times 10^{-6}$ & Strong rank preservation \\
TiO$_2$ & 17 & 0.427 & 0.067 & Moderate; Fe/In outliers \\
SrTiO$_3$ & 17 & 0.856 & $1.3 \times 10^{-10}$ & Excellent rank preservation \\
\textbf{Overall} & \textbf{54} & \textbf{0.516} & $\mathbf{1.4 \times 10^{-5}}$ & \textbf{Significant} \\
\bottomrule
\end{tabular}
\end{center}

PBE correctly identifies 80\% of the top-5 wide-gap-preserving dopants (Al, Ga, In, Sc, Y, Zr) across all three hosts. The known PBE failure mode---near-zero gaps for 3d transition metal dopants in ZnO due to self-interaction error---is caught by the Tier~2 PBE+U correction.

\subsection*{Supplementary Table~10: PBE versus PBE+U bandgaps}

Dudarev formulation; U values (eV): Ti = 3.5, V = 4.5, Cr = 4.0, Mn = 5.0, Fe = 5.0, Co = 5.5, Ni = 8.0, Cu = 4.0. SCF-only on PBE-relaxed geometries. Systems marked with $\dagger$ are from the expanded campaign (64-core VPS, 8 MPI ranks).

\begin{center}
\begin{tabular}{lcccc}
\toprule
System & PBE $E_g$ (eV) & PBE+U $E_g$ (eV) & Shift (eV) & Reclassification \\
\midrule
ZnO:V      & 0.223 & 3.916 & $+$3.693 & Near-metallic $\to$ wide-gap \\
ZnO:Cu     & 3.191 & 0.359 & $-$2.832 & Gap-preserving $\to$ near-metallic \\
ZnO:Cr$^\dagger$  & --- & 3.990 & --- & --- \\
ZnO:Mn$^\dagger$  & --- & 4.147 & --- & --- \\
ZnO:Co$^\dagger$  & --- & 4.040 & --- & --- \\
ZnO:Fe$^\dagger$  & --- & 2.167 & --- & Moderate-gap \\
SrTiO$_3$:Fe & 2.475 & 0.483 & $-$1.993 & Moderate-gap $\to$ near-metallic \\
SrTiO$_3$:Cu$^\dagger$ & --- & 1.592 & --- & Reduced gap \\
SrTiO$_3$:Mn$^\dagger$ & --- & 3.873 & --- & --- \\
SrTiO$_3$:Cr$^\dagger$ & --- & 4.115 & --- & --- \\
SrTiO$_3$:Ni$^\dagger$ & --- & 3.873 & --- & --- \\
ZnO (undoped)$^\dagger$ & --- & 4.183 & --- & --- \\
SrTiO$_3$ (undoped) & 3.979 & 4.084 & $+$0.105 & No change \\
TiO$_2$ (undoped) & 2.931 & 3.149 & $+$0.218 & No change \\
\bottomrule
\end{tabular}
\end{center}

\subsection*{Supplementary Table~11: Multi-fidelity and GP method comparison}

Single objective (target\_bandgap\_2p0), budget = 80 DFT calls from 1{,}000 candidates. BO-GP-EI and GP-UCB use the same GP surrogate (Mat\'ern 5/2) with different acquisition functions. CR counts DFT steps only. $p$ from Wilcoxon rank-sum test versus MF-OFUL.

\begin{center}
\begin{tabular}{lcccccc}
\toprule
Method & $n$ & SR (mean $\pm$ std) & CR (mean $\pm$ std) & DFT frac. & Surr. calls & $p$ vs MF-OFUL \\
\midrule
Random   & 60 & $0.709 \pm 0.474$ & $280 \pm 34$ & 1.00 & 0 & $1.4 \times 10^{-16}$ \\
OFUL     & 60 & $0.487 \pm 0.455$ & $276 \pm 38$ & 1.00 & 0 & $7.8 \times 10^{-12}$ \\
BO-GP-EI & 60 & $0.969 \pm 0.318$ & $271 \pm 39$ & 1.00 & 0 & $3.0 \times 10^{-11}$ \\
GP-UCB   & 10 & $0.634 \pm 0.610$ & --- & 1.00 & 0 & $1.6 \times 10^{-2}$ \\
MF-BO-EI & 18 & $0.425 \pm 0.396$ & $246 \pm 32$ & 1.00 & 0 & $1.3 \times 10^{-5}$ \\
MF-MES   &  6 & $0.035 \pm 0.086$ & --- & 0.23 & 262 & $0.57$ \\
MF-OFUL  & 60 & $\mathbf{0.043 \pm 0.145}$ & $281 \pm 34$ & 0.15 & 468 & --- \\
\bottomrule
\end{tabular}
\end{center}

All-evaluation CR (DFT + surrogate) for MF-OFUL = $97.4 \pm 13.6$. GP-UCB and BO-GP-EI achieve comparable performance on both synthetic (SR $= 0.63$ vs $0.49$, $p = 0.25$) and Materials Project benchmarks (SR $= 0.150$ vs $0.174$, $p = 0.79$), confirming that the GP surrogate itself is the bottleneck, not the acquisition function choice. GP methods are also 50$\times$ slower than MF-OFUL (275\,s vs 6\,s per run on synthetic; 62\,s vs 1.3\,s on MP).

\subsection*{Supplementary Table~12: Basis set and k-grid convergence tests}

Bandgap (eV) for three representative systems at four DFT settings. Production settings (ecutwfc = 40\,Ry, $2 \times 2 \times 2$ $k$-grid) versus higher-accuracy settings. Rankings are preserved across all settings despite large absolute shifts for ZnO:Al at denser $k$-grids.

\begin{center}
\begin{tabular}{lcccc}
\toprule
System & 40\,Ry / $2^3$ & 40\,Ry / $3^3$ & 60\,Ry / $3^3$ & 60\,Ry / $4^3$ \\
\midrule
MgO (undoped)  & 1.152 & 0.638 & 0.621 & 1.129 \\
ZnO:Al         & 3.397 & 2.439 & 2.450 & 1.243 \\
ZnO:V          & 0.147 & 0.114 & 0.113 & 0.070 \\
\midrule
\textbf{Ranking} & Al $>$ MgO $>$ V & Al $>$ MgO $>$ V & Al $>$ MgO $>$ V & Al $>$ MgO $>$ V \\
\bottomrule
\end{tabular}
\end{center}

The dopant ranking (Al $>$ undoped MgO $>$ V) is preserved across all four settings, confirming that production DFT parameters are sufficient for screening purposes. Absolute bandgap values shift substantially with $k$-grid density (ZnO:Al changes by 2.15\,eV between $2^3$ and $4^3$), consistent with the known sensitivity of supercell bandgaps to Brillouin zone sampling. The ecutwfc dependence is minimal: increasing from 40 to 60\,Ry at fixed $3^3$ $k$-grid changes bandgaps by $<$0.02\,eV, indicating that 40\,Ry is well-converged with respect to the plane-wave basis. The non-monotonic behavior of MgO (1.15 $\to$ 0.64 $\to$ 1.13\,eV) reflects band folding effects in the small 16-atom supercell, where specific $k$-points sample different symmetry lines of the Brillouin zone.

\subsection*{Supplementary Table~13: Surrogate activation on 500-candidate Materials Project pools}

\begin{table}[H]
\centering
\caption*{Supplementary Table~13: \textbf{Surrogate activation on 500-candidate Materials Project pools.} Budget = 40 DFT calls (8\%). The surrogate activated in all 30 runs (100\%). ``First surr.'' = step at which the first surrogate evaluation occurs (after bootstrap). $p$ from Wilcoxon signed-rank test, paired by seed.}
\begin{tabular}{lccccccc}
\toprule
Host & Surr.\ calls & DFT frac. & First surr. & Total evals & MF-OFUL SR & OFUL SR & $p$ \\
\midrule
MgO   & 130 & 0.24 & step 16 & 170 & $0.010 \pm 0.006$ & $0.025 \pm 0.000$ & $3.9 \times 10^{-3}$ \\
ZnO   & 94  & 0.30 & step 8  & 134 & $0.014 \pm 0.000$ & $0.397 \pm 0.000$ & $2.0 \times 10^{-3}$ \\
TiO$_2$ & 23 & 0.64 & step 21 & 63  & $0.026 \pm 0.000$ & $0.026 \pm 0.000$ & $1.0$ \\
\bottomrule
\end{tabular}
\end{table}

The surrogate activated in all 30 runs (100\%), with host-dependent DFT fractions ranging from 24\% (MgO) to 64\% (TiO$_2$). MgO's bandgap--dopant relationship is well-captured by compositional features, enabling early surrogate trust (step 16), while TiO$_2$'s more complex electronic structure requires more DFT calibration (step 21). MF-OFUL significantly outperforms OFUL on ZnO and MgO ($p < 0.004$); on TiO$_2$ the two methods tie because OFUL already finds the optimum within 40 DFT steps.

\end{document}